\documentclass[12pt]{article}
\usepackage[utf8]{inputenc}
\usepackage[T1]{fontenc}
\usepackage{lmodern}
\usepackage[colorlinks,citecolor=blue,urlcolor=blue]{hyperref}
\usepackage{amsthm,amsmath,amsfonts,amssymb}
\usepackage{subcaption,graphicx,enumerate,setspace}
\usepackage{booktabs, array}
\usepackage[parfill]{parskip}
\usepackage[round,comma]{natbib}
\usepackage{algorithm,algpseudocode}
\newcommand{\pp}[2]{\mathtt{#1}_{\mathrm{#2}}}

\usepackage[a4paper,hscale=0.8,vscale=0.8]{geometry}

\begin{document}
\title{%
Simulation-Based Fitting of Intractable Models via Sequential Sampling
and Local Smoothing 
}
\author{%
Guido Masarotto \\
Department of Statistical Sciences\\
University of Padova, ITALY
}

\maketitle

\begin{abstract} 

This paper presents a comprehensive algorithm for fitting generative
models whose likelihood, moments, and other quantities typically used
for inference are not analytically or numerically tractable.  The
proposed method aims to provide a general solution that requires only
limited prior information on the model parameters.  The algorithm
combines a global search phase, aimed at identifying the region of the
solution, with a local search phase that mimics a trust region version
of the Fisher scoring algorithm for computing a quasi-likelihood
estimator.  Comparisons with alternative methods demonstrate the strong
performance of the proposed approach.  An R package implementing the
algorithm is available on CRAN.


\end{abstract}

\section{Introduction}
\label{sec:int}

Across many fields, researchers frequently encounter parametric models
whose likelihoods and other inference-related quantities are
analytically intractable or computationally prohibitive; yet from which data can
still be simulated. For at least five decades, simulation-based
techniques for estimating such complex models have been employed in economics and finance
\citep[see, e.g,][]{McFadden1989, Pakes1989, Gourieroux1997, Mariano2000, Forneron2018}.
Following the introduction of the Approximate Bayesian
Computation (ABC) framework \citep[e.g.,][]{Sisson2018}, these methods have
proliferated and are now used in diverse disciplines including biology
\citep{Tavare1997, Pritchard1999, Beaumont2002, Beaumont2010,
Marchand2017}, hydrology \citep{Hull2024}, and astronomy
\citep{Cameron2012, Alvey2024, Moser2024, Barret2024, Dupourque2025,
Fischbacher2025}.

While the past decades have seen extensive research into
simulation-based Bayesian approaches such as ABC, the development of
general-purpose algorithms for the frequentist setting has been less
pronounced. Despite notable exceptions \citep[e.g.,][]{Cox2012},
a gap persists for robust and broadly applicable solutions. 
This paper aims to help fill this gap by proposing 
a new comprehensive algorithm, which we refer to as \texttt{ifit}, for
fitting a generative model $ \bigl\{p_{\boldsymbol{\vartheta}} :
\boldsymbol{\vartheta} \in \Theta \bigr\} $ to observed data
$\mathbf{y}_{\mathrm{obs}}$.  The algorithm operates under the following
assumptions: 
\begin{enumerate}[(i)] 
    \item 
        It is possible to simulate
    pseudo-data $\mathbf{y} \sim p_{\boldsymbol{\vartheta}}$ for every
    $\boldsymbol{\vartheta} \in \Theta$.  
    \item 
        The model parameter
    $\boldsymbol{\vartheta}$ is a $p$-dimensional real vector. The only
    prior information available about $\boldsymbol{\vartheta}$ takes the
    form of \emph{boundary constraints}, whereby each component
$\vartheta_i$ is known to lie within a specific interval $[l_i, u_i]$.
\end{enumerate}

The considered estimator relies on selecting $q \ge p$ features,
$\mathbf{t}_{\mathrm{obs}} = \Psi(\mathbf{y}_{\mathrm{obs}})$, intended
to capture the information in the data. The objective is to find the
parameter vector $\widehat{\boldsymbol{\vartheta}}$ that minimizes the
distance between $\mathbf{t}_{\mathrm{obs}}$ and
$E_{\widehat{\boldsymbol{\vartheta}}}\bigl\{ \Psi(\mathbf{y})\bigr\}$.
Thus, the estimator belongs to the class of methods that
\citet{Jiang2004} refer to as \emph{indirect inference based on
intermediate statistics}. This estimator can also be interpreted as an
application of the \emph{simulated generalized method of moments}
\citep{McFadden1989, Pakes1989}. In addition, it borrows key ideas ---
such as sequential exploration and local surface approximation --- from
the \emph{response surface methodology} (RSM) and \emph{surrogate-based
optimization} (SBO) frameworks \citep{Box1951, Myers2016, Jones2001}.

The algorithm begins with a global search to identify a promising
starting point, followed by a local search phase that refines the
solution using a trust region version of the Fisher scoring algorithm
for computing a quasi-likelihood estimator. Since the moments and other
quantities associated with the intermediate statistic $\mathbf{t} =
\Psi(\mathbf{y})$ are unknown, they are approximated via simulation
using local regression techniques.  To reduce the number of simulations,
the sampling scheme is sequential.  After iteration $k$ of the
algorithm, $N_k$ pairs $[\boldsymbol{\vartheta}_i, \mathbf{t}_i]$, with
$\mathbf{t}_i = \Psi(\mathbf{y}_i)$ and $\mathbf{y}_i \sim
p_{\boldsymbol{\vartheta}_i}$, are available. These pairs are then used
to determine the location of the next set of simulations, i.e., to
select $\boldsymbol{\vartheta}_{N_k+1}, \ldots,
\boldsymbol{\vartheta}_{N_{k+1}}$, with the objective of generating new
parameter samples that are progressively closer to the final estimate.

The \texttt{ifit} algorithm provides a reliable and computationally
efficient solution. In particular, the combination of sequential
sampling and local regression contrasts with existing implementations of
simulation-based indirect inference or the generalized method of moments
(see, e.g., \citealp{Gourieroux1997} and \citealp{Cox2012}) which often lack
mechanisms for robust global exploration --- an aspect that is crucial
when prior information on the parameters is weak --- and estimate the
required quantities (e.g., the mean of the intermediate statistic)
through costly repeated simulations for a fixed
$\boldsymbol{\vartheta}$. By contrast, like RSM and SBO, \texttt{ifit}
seeks to minimize the computational cost by exploiting local regression
to extract information from the entire simulated dataset. Moreover,
unlike other implementations, which require the user to manually select
the components of the step-size vector used for numerical
differentiation --- a choice that, for example, \citet{Cox2012}
explicitly acknowledges as difficult and largely heuristic ---
\texttt{ifit} provides an automatic and data-driven mechanism for
determining a suitable step-size. Furthermore, as shown in Section
\ref{sec:num}, \texttt{ifit} can be more efficient than some ABC
algorithms, at least under diffuse prior settings, achieving comparable
or even better accuracy while using substantially fewer computational
resources.

The remainder of the paper is organized as follows.
Section~\ref{sec:back} provides background and contextual information,
assuming that the moments of the intermediate statistic are analytically
available. Section~\ref{sec:alg} presents the \texttt{ifit} algorithm in
detail.  Section~\ref{sec:toad} illustrates the application of the
procedure to a real dataset.  Section~\ref{sec:num} reports the results
of an extensive Monte Carlo. Finally, Section~\ref{sec:conc} concludes
with a discussion of the findings and potential directions for future
research.

\section{Background and notation}
\label{sec:back}

Let $\mathbf{y}_{\mathrm{obs}}$ denote the observed data, and let
$\{p_{\boldsymbol{\vartheta}} : \boldsymbol{\vartheta} \in \Theta\}$
be the parametric model used for interpretation.
The parameter space is a $p$-dimensional hypercube,
$$
    \Theta = \bigl\{(\vartheta_1, \ldots, \vartheta_p)^{\mathrm{T}} \in
   \mathbb{R}^p : l_i \le \vartheta_i \le u_i\bigr\}.
$$
The data structure may be arbitrary, but for asymptotic arguments
we assume that the sample size is indexed by an integer $n$.
In addition, let $\mathbf{t}_{\mathrm{obs}} = \Psi(\mathbf{y}_{\mathrm{obs}}) \in \mathbb{R}^q$
be the chosen summary statistics, and let
$\tau(\boldsymbol{\vartheta}) = E_{\boldsymbol{\vartheta}}\{\Psi(\mathbf{y})\}$
denote their expectation.

The indirect inference estimator based on the bridge
relationship $\tau(\cdot)$ \citep{Jiang2004},
which is equivalently the generalized method of moments estimator
\citep{Hansen1982,Gourieroux1997}, is defined as
$$
\widehat{\boldsymbol{\vartheta}}
= \arg\min_{\boldsymbol{\vartheta}}
\|\mathbf{t}_{\mathrm{obs}} - \tau(\boldsymbol{\vartheta})\|_{\mathbf{V}}^2,
$$
where $\mathbf{V}$ is a symmetric positive-definite $q\times q$
weighting matrix, and
$\|\cdot\|_{\mathbf{V}}$ denotes the corresponding Mahalanobis norm,
i.e., for any $\mathbf{x} \in \mathbb{R}^q$,
$$
\|\mathbf{x}\|_{\mathbf{V}} =
\bigl(\mathbf{x}^{\mathrm{T}}\mathbf{V}^{-1}\mathbf{x}\bigr)^{1/2}.
$$

If the model is correctly specified,
i.e., $\mathbf{y}_{\mathrm{obs}} \sim p_{\boldsymbol{\vartheta}_0}$ for some
$\boldsymbol{\vartheta}_0 \in \Theta$,
then under mild conditions
$\widehat{\boldsymbol{\vartheta}}$ is consistent for any $\mathbf{V}$.
The main requirements are that  
(i)~$\tau(\cdot)$ identifies the parameter, that is,
$\tau(\boldsymbol{\vartheta}') \ne \tau(\boldsymbol{\vartheta}'')$
whenever $\boldsymbol{\vartheta}' \ne \boldsymbol{\vartheta}''$, and  
(ii)~$\mathbf{t}_{\mathrm{obs}}-\tau(\boldsymbol{\vartheta}_0)$
vanishes asymptotically, at least in probability.

Furthermore, under the usual regularity conditions,
a standard first-order expansion shows that if
$
\sqrt{n}\bigl\{\mathbf{t}_{\mathrm{obs}}-\tau(\boldsymbol{\vartheta}_0)\bigr\}
$
is asymptotically normal with mean zero and
dispersion matrix $\boldsymbol{\Sigma}(\boldsymbol{\vartheta}_0)$, then
$$
\sqrt{n}\bigl(\widehat{\boldsymbol{\vartheta}}-\boldsymbol{\vartheta}_0\bigr)
\stackrel{D}{\longrightarrow}
N_p\!\left(
    \mathbf{0}_p,\, \boldsymbol{\Omega}^{-1}(\boldsymbol{\vartheta}_0)
\right),
$$
where
$$
\boldsymbol{\Omega}(\boldsymbol{\vartheta})
=
\mathbf{J}^{\mathrm{T}}(\boldsymbol{\vartheta})\mathbf{V}^{-1}\mathbf{J}(\boldsymbol{\vartheta})
\bigl\{
\mathbf{J}^{\mathrm{T}}(\boldsymbol{\vartheta})\mathbf{V}^{-1}
\boldsymbol{\Sigma}(\boldsymbol{\vartheta})
\mathbf{V}^{-1}\mathbf{J}(\boldsymbol{\vartheta})
\bigr\}^{-1}
\mathbf{J}^{\mathrm{T}}(\boldsymbol{\vartheta})\mathbf{V}^{-1}\mathbf{J}(\boldsymbol{\vartheta}),
$$
and
$$
\mathbf{J}\bigl(\boldsymbol{\vartheta}\bigr)
=
\dfrac{\partial \tau(\boldsymbol{\vartheta})}
{\partial \boldsymbol{\vartheta}^{\mathrm{T}}}.
$$
While the asymptotic normality of the estimator depends crucially on that of the intermediate statistics, 
the expression for the asymptotic covariance matrix derived above continues to hold 
as long as the asymptotic variance of the intermediate statistics exists.
It can also be shown that the asymptotic variance of the
estimator is minimized by setting
$\mathbf{V}=\boldsymbol{\Sigma}(\boldsymbol{\vartheta}_0)$,
in which case
\begin{equation}
\boldsymbol{\Omega}(\boldsymbol{\vartheta})=
\mathbf{J}^{\mathrm{T}}(\boldsymbol{\vartheta})
\boldsymbol{\Sigma}^{-1}(\boldsymbol{\vartheta})
\mathbf{J}(\boldsymbol{\vartheta}).
\label{eqn:asyvar}
\end{equation}

Since $\boldsymbol{\vartheta}_0$ is unknown, direct use
of the optimal weighting matrix is not possible. Consequently,
two-step or multi-step strategies have been proposed,
where an initial consistent but inefficient estimate of
$\boldsymbol{\vartheta}_0$ is used to construct a
consistent estimate of the optimal weighting matrix
$\boldsymbol{\Sigma}(\boldsymbol{\vartheta}_0)$
\citep[see, e.g.,][]{Hansen1996}.
In this paper, however, I follow
a slightly different approach, defining the desired estimator as the
solution to the equation 
\begin{equation}
g(\widehat{\boldsymbol{\vartheta}}) = \mathbf{0}_p
\label{eqn:esteq}
\end{equation}  
where the estimating function is defined as
\begin{equation}
g\bigl(\boldsymbol{\vartheta}\bigr)    
=
\mathbf{J}^{\mathrm{T}}\bigl(\boldsymbol{\vartheta}\bigr)
\boldsymbol{\Sigma}^{-1}\bigl(\boldsymbol{\vartheta}\bigr)
\bigl\{\mathbf{t}_{\mathrm{obs}}-\tau(\boldsymbol{\vartheta})\bigr\}.
\label{eqn:grad}
\end{equation}
Assuming the necessary regularity conditions hold,
it is straightforward to show that any solution to \eqref{eqn:esteq}
has the same asymptotic distribution as the optimal indirect estimator.
Note the similarities between \eqref{eqn:grad} and the quasi-likelihood 
score \citep{Heyde1997}.

When the number of features used to summarize the data $q$ exceeds the
number of parameters $p$, the estimation results can also be employed to
assess the adequacy of the chosen model—or, more precisely, the model’s
ability to reproduce the mean of the summary statistics.  In particular,
the Sargan–Hansen test statistic 
\begin{equation} 
    \mathrm{SH} = \left\|
    \mathbf{t}_{\mathrm{obs}}-\tau\bigl(\widehat{\boldsymbol\vartheta}\bigr)
    \right\|^2_{\boldsymbol{\Sigma}(\widehat{\boldsymbol\vartheta})},
\label{eqn:jtest} 
\end{equation}  
is asymptotically distributed as a
$\chi^2_{q-p}$ variable when the model is correctly specified and
$\mathbf{t}_{\mathrm{obs}}$ is asymptotically normal \citep{Hansen1982}.
Moreover, it can be useful to inspect the individual standardized
components of the summary statistics, namely, 
\begin{equation}
\bigl\{\mathrm{diag}\boldsymbol{\Sigma}(\widehat{\boldsymbol{\vartheta}})\bigr\}^{-1/2}
\bigl\{\mathbf{t}_{\mathrm{obs}}-\tau(\widehat{\boldsymbol{\vartheta}})\bigr\}
\label{eqn:sscore} 
\end{equation} 
to identify specific shortcomings of the selected model. 
In \eqref{eqn:sscore}, $\{\mathrm{diag}\, \mathbf{S}\}^{-1/2}$ 
denotes a diagonal matrix with elements $1/\sqrt{s_{ii}}$, where 
$s_{ii}$ are the diagonal elements of $\mathbf{S}$.

\section{The \texttt{ifit} algorithm}
\label{sec:alg}

\subsection{Preliminary remarks}

\paragraph{Tunable parameters.} 
The proposed algorithm depends on several constants, listed in Table
\ref{tab:const}, that users can select to tune specific aspects of its
behavior. While the number of tunable parameters is relatively large to
enhance flexibility, the default settings, also provided in Table
\ref{tab:const}, have proven effective in a wide variety of
applications. For example, these settings will be used without
modification for all the models considered in Section~\ref{sec:num}.

\paragraph{Scaling of the parameters.}
As part of its effort to remain robust
across different models, \texttt{ifit} attempts to automatically scale the
parameters. This is, for example, the purpose of the matrices
$\mathbf{D}_{\mathrm{something}}$ used 
to quantify distances between parameter vectors
(see Algorithm~\ref{alg:glo}, step~\ref{alg:ginit}, and 
Algorithm~\ref{alg:local}, step \ref{alg:lest}). 
The same rationale motivates the shape of the
trust region used during the local search phase. 
It is a hypercube, since this choice allows each side of
the region to be rescaled according to the size of the corresponding
parameter (Algorithm~\ref{alg:local}, Step~\ref{alg:delta}). However, despite this built-in
mechanism, as with most optimization procedures, ifit probably benefits
from a preliminary reparameterization of the model so that all
parameters have roughly comparable magnitudes.

\begin{table}[tp]
\centering
\begin{tabular}{>{\raggedright}p{0.33\textwidth}p{0.60\textwidth}}
\toprule
Constants (default values in brackets) & Description \\
\midrule
$\pp{N}{init} (1000)$ & 
Size of the initial random Latin hypercube
sample (Algorithm~\ref{alg:glo}, step~\ref{alg:ginit})\\
$\pp{N}{elite} (100)$, $\pp{A}{elite} (0.5)$ & 
Parameters controlling the size of the elite sample (Algorithm~\ref{alg:glo}, step~\ref{alg:elite})\\
$\pp{Tol}{global} (0.1)$, $\pp{Tol}{local} (1)$,
$\pp{Tol}{model} (1.5)$ &
Tolerances. The first two are used in the stopping criteria of the
global and local phases, respectively
(Algorithm~\ref{alg:glo}, step~\ref{alg:gstop} and
Algorithm~\ref{alg:local}, step~\ref{alg:lstop});
the latter is used to check the adequacy of the current local
linear model (Algorithm~\ref{alg:local}, step~\ref{alg:mok})\\
$\pp{NFit}{local} (4000)$ & 
Desired size of the neighborhood used to fit
the local linear model. During the local search, the neighborhood size grows
progressively from $\pp{N}{elite}$ up to this limit
(Algorithm~\ref{alg:local}, steps~\ref{alg:lstop} and~\ref{alg:lupd})\\
$\pp{NAdd}{global} (100)$, $\pp{NAdd}{local} (10)$ &
Number of new simulations performed at each iteration during the global and
local phases, respectively (Algorithm~\ref{alg:glo}, step~\ref{alg:gadd}, and
Algorithm~\ref{alg:local}, step~\ref{alg:ladd})\\
$\pp{Rho}{max} (0.1)$ & 
Maximum radius of the trust region
(Algorithm~\ref{alg:local}, steps~\ref{alg:linit},~\ref{alg:delta}, and~\ref{alg:mok})\\
$\pp{Lambda}{} (0.1)$ & 
Constant used to smooth the Jacobian and
covariance estimates (Algorithm~\ref{alg:local}, step~\ref{alg:lest})\\
\bottomrule
\end{tabular}
\caption{%
User-selectable \texttt{ifit} constants and their default values.
}
\label{tab:const}
\end{table}

\subsection{The first phase: global search}
\label{subsec:algglo}

\begin{algorithm}[htp]
    \caption{The global search phase of the \texttt{ifit} algorithm}
    \label{alg:glo}
\begin{algorithmic}[1]
    \State \label{alg:ginit} 
    \emph{Global search initialization}. Set $k \gets 0$ and $N_0 \gets
    \pp{N}{init}$. Let $\mathbf{D}_0$ be the $p\times p$
    diagonal matrix with elements $(u_i-l_i)^2$ on its diagonal. Generate $\boldsymbol{\vartheta}_1, \ldots, \boldsymbol{\vartheta}_{N_{0}}$ within the parameter space $\Theta$ using Latin hypercube sampling \citep{McKay1979}. Simulate the corresponding summary statistics $\mathbf{t}_1, \ldots, \mathbf{t}_{N_0}$.

    \State \label{glob:begin} \emph{Bridge function estimation}.
    For $i=1, \ldots, N_k$, calculate
    $$
    \widehat{\boldsymbol{\tau}}_{k, i} =
    \sum_{\{r:
        \|\boldsymbol{\vartheta}_r-\boldsymbol{\vartheta}_i\|_{\mathbf{D}_0}
        \le
        \overline{d}_{k,i}\}}
    \left\{1-
        \left(\dfrac{
            \|\boldsymbol{\vartheta}_r-\boldsymbol{\vartheta}_i\|_{\mathbf{D}_0}}
            {\overline{d}_{k,i}}
    \right)^3\right\}^3 \mathbf{t}_r,
    $$
    where $\overline{d}_{k,i}$ is chosen so that the neighborhood size
    equals $\lceil N_k^{1/2}\rceil$ ($\lceil x \rceil$ denotes the
    smallest integer greater than or equal to $x$).

    \State \label{alg:gweight} 
    \emph{Weighting matrix}. Set $\boldsymbol{\Sigma}_k \gets \mathbf{S}_k \mathbf{R}_k \mathbf{S}_k$, where $\mathbf{S}_k$ is the $q \times q$ diagonal matrix with the median absolute deviations (MADs) of the components of the residual vectors $(\mathbf{t}_i-\widehat{\boldsymbol{\tau}}_{k,i})$ for $i=1, \ldots, N_k$ on its diagonal, and $\mathbf{R}_k$ is the correlation matrix of the Gaussian scores of the same residual vectors \citep{Boudt2012}.

    \State \label{alg:elite} 
    \emph{Elite sample selection}. Set
    $
    E_k \gets \left\lceil \pp{N}{elite} + (\pp{N}{init} - \pp{N}{elite}) \pp{A}{elite}^{(N_k/\pp{N}{init})^2} \right\rceil.
    $
    Then identify the $E_k$ parameter vectors
    $\boldsymbol{\vartheta}_{i_{k,1}}, \ldots,
    \boldsymbol{\vartheta}_{i_{k,E_k}}$ corresponding to the $E_k$
    smallest Mahalanobis distances
    $\|\mathbf{t}_{\mathrm{obs}}-\widehat{\boldsymbol{\tau}}_{k,
    i}\|_{\boldsymbol{\Sigma}_k}$.

    \State \label{alg:gstop}
    \emph{Convergence check}. Proceed to the local search phase (Algorithm~\ref{alg:local}) if, for each parameter $i=1, \ldots, p$,
    $$
    sd_{k, i} < \max\bigl(1, |m_{k, i}|\bigr) \cdot \pp{Tol}{global},
    $$
    where $m_{k,i}$ and $sd_{k,i}$ are the sample mean and standard deviation of the $i$th component of the parameter vectors in the elite sample.

    \State \label{alg:gadd} 
    \emph{Elite sample reproduction}. Set $N_{k+1} \gets N_k+\pp{NAdd}{global}$ and let $\mathbf{C}_k$ be the sample covariance matrix of the elite sample. Sample $\boldsymbol{\vartheta}_{N_k+1}, \ldots, \boldsymbol{\vartheta}_{N_{k+1}}$ from a mixture of $E_k$ truncated multivariate normal distributions with support on $\Theta$. Each component of the mixture is centered at a mean vector drawn from the elite sample $\{\boldsymbol{\vartheta}_{i_{k,1}}, \ldots, \boldsymbol{\vartheta}_{i_{k, E_k}}\}$ and has covariance matrix $\mathbf{C}_k$. Simulate the corresponding summary statistics $\mathbf{t}_{N_k+1}, \ldots, \mathbf{t}_{N_{k+1}}$.
    \State \emph{Update and iterate.}  Set $k \gets k+1$ and return to step~\ref{glob:begin}.
    \algstore{brbreak}
\end{algorithmic}
\end{algorithm}

The purpose of the global search phase, detailed in
Algorithm~\ref{alg:glo}, is to select a set of points in the parameter
space $\Theta$ --- at least $\pp{N}{elite}$, using the notation
introduced in the algorithm and summarized in Table~\ref{tab:const} --- that
are close to the desired solution. The approach is common to global
search strategies that operate with a population of candidate solutions
(e.g., genetic algorithms, cross-entropy optimization, and other
stochastic search methods; see, e.g., \citealp{EibenSmith2015,
RubinsteinKroese2004, Spall2003}).  At any given iteration, a population
of particles is available. These particles are ranked, and only the best
ones, referred to as the elite sample (step~\ref{alg:elite} in the
algorithm), are allowed to generate descendants that are added to the
population (step~\ref{alg:gadd}). The algorithm terminates — and in this
case, proceeds to the second phase, the local search, see
Subsection~\ref{subsec:algloc} — when the elite sample is sufficiently
concentrated (step~\ref{alg:gstop}). The ranking of the particles is
based on an approximation of the optimal objective function for indirect
inference 
$$
\|\mathbf{t}_{\mathrm{obs}}-\tau(\boldsymbol{\vartheta})\|^2_{\boldsymbol{\Sigma}(\boldsymbol{\vartheta}_0)}.
$$ 
Specifically, the bridge function $\tau(\cdot)$ is estimated using
k-nearest neighbors regression (step~\ref{glob:begin}), while the
weighting matrix is approximated under the assumption that both the
variances and the correlations of the summary statistics do not depend
on $\boldsymbol\vartheta$ (step~\ref{alg:gweight}). 
However, since the estimator used is robust (combining the median
absolute deviations with the correlation coefficients of the Gaussian
scores), the impact of this simplifying homoscedasticity assumption is
reduced, especially in the later iterations when many points have
hopefully been sampled near the target solution. In any case, 
the consistency of the estimatori --- the only
crucial property in this global search phase --- does not depend on the
particular weighting matrix employed.

A review of Table~\ref{tab:const} and step~\ref{alg:elite} of the algorithm
shows that, in the default setting where $\pp{A}{elite}=0.5$,
the initial size of the elite sample is
over 50\% of the population size. This, combined with the fact
that the mixture from which the offspring are generated has a
covariance matrix equal to twice that of the elite
sample (step~\ref{alg:gadd}), ensures a good exploratory capability
in the initial stages. 
Progressively, the size of the elite sample decreases 
and eventually equals $\pp{N}{elite}$, whose default value is 100, 
thereby reducing the elite-to-population ratio. 
The rate of this decrease depends on $\pp{A}{elite}$.

A distinctive feature of the algorithm is that, unlike other
population-based optimization methods, in \texttt{ifit} new
offspring are added to rather than replacing the existing ones. This is
because the objective function is estimated
using local regression, and thus the algorithm aims to
leverage all the information gathered. For the same reason, the
default number of new offspring generated at each iteration
($\pp{NAdd}{global}=100$) is relatively small. The idea is that
this allows for a more gradual learning of the objective function, so that
large initial estimation errors can be corrected before they have a real
impact.

\subsection{The second phase: local search}
\label{subsec:algloc}

\begin{algorithm}[htp]
    \caption{The local search phase of the \texttt{ifit} algorithm}
    \label{alg:local}
\begin{algorithmic}[1]
    \algrestore{brbreak}
    \State \label{alg:linit} \emph{Local search initialization}.
    Set $L_k \gets \pp{N}{elite}$ and $\rho_k \gets \pp{Rho}{max} / 10$, 
    $\widehat{\boldsymbol{\vartheta}}_k \gets \boldsymbol{\vartheta}_{\textit{best}}$,
    where the subscript \textit{best} denotes the index of the point found during the global phase (Algorithm~\ref{alg:glo}) that minimizes the Mahalanobis distance:
    $$
    \|\mathbf{t}_{\text{obs}}-\widehat{\boldsymbol{\tau}}_{k,\textit{best}}\|
    _{\boldsymbol{\Sigma}_k}
    = \min_{i=1, \ldots, N_k}
    \|\mathbf{t}_{\text{obs}}-\widehat{\boldsymbol{\tau}}_{k, i}\|
    _{\boldsymbol{\Sigma}_k}.
    $$

    \State \label{alg:lest} 
    \emph{Bridge function, Jacobian, and covariance estimation}.
    Let $\mathbf{D}_k$ be the $p\times p$ diagonal matrix with $\max(1, \widehat\vartheta_{k,i}^2)$ on its diagonal, where $\widehat\vartheta_{k,i}$ is the $i$th component of $\widehat{\boldsymbol\vartheta}_k$.
    Based on the $L_k$ pairs $[\boldsymbol\vartheta_i, \mathbf{t}_i]$
    closest to $\widehat{\boldsymbol\vartheta}_k$ (in terms of the
    distance
    $\|\boldsymbol\vartheta-\widehat{\boldsymbol\vartheta}_k\|_{\mathbf{D}_k}$), fit the multivariate linear regression model
    $$
    \mathbf{t}_i = \boldsymbol{\tau}+\mathbf{B}(\boldsymbol\vartheta_i-\widehat{\boldsymbol\vartheta}_k) +\mathbf{Error}_i.
    $$
    Let $\widehat{\boldsymbol\tau}_k, \widehat{\mathbf{B}}_k, \widehat{\mathbf{W}}_k$, and $\widehat{\mathbf{H}}_k$ be the least-squares estimates of $\boldsymbol\tau$, $\mathbf{B}$, the error covariance $\mathrm{Var}(\mathbf{Error}_i)$, and $\mathrm{Var}(\widehat{\boldsymbol\tau}_k)$, respectively.
    Update the estimates as follows:
    \begin{align*}
        \widehat{\mathbf{J}}_k &\gets \begin{cases}
            \widehat{\mathbf{B}}_k & \text{if } L_k=\pp{N}{elite}, \\
            (1-\pp{Lambda}{})\widehat{\mathbf{J}}_{k-1}+\pp{Lambda}{} \widehat{\mathbf{B}}_k & \text{otherwise},
    \end{cases}\\
        \widehat{\boldsymbol\Sigma}_k &\gets \begin{cases}
            \widehat{\mathbf{W}}_k & \text{if } L_k=\pp{N}{elite}, \\
            (1-\pp{Lambda}{})\widehat{\boldsymbol{\Sigma}}_{k-1}+\pp{Lambda}{} \widehat{\mathbf{W}}_k & \text{otherwise},
    \end{cases}\\
            \widehat{\mathbf{g}}_k &\gets \widehat{\mathbf{J}}_k^{\mathrm{T}}
        \widehat{\boldsymbol\Sigma}_k^{-1}\bigl(\mathbf{t}_{\text{obs}}-\widehat{\boldsymbol\tau}_k\bigr),
        \text{ and }
        \widehat{\boldsymbol\Omega}_k \gets 
        \widehat{\mathbf{J}}_k^{\mathrm{T}}
        \widehat{\boldsymbol\Sigma}_k^{-1}
        \widehat{\mathbf{J}}_k
    \end{align*}

    \State \label{alg:delta} 
    \emph{Candidate point}.
    Compute the update step $\boldsymbol\delta_k=(\delta_1, \ldots, \delta_p)^{\mathrm{T}}$ by solving the linear programming problem:
    $$
    \min_{\boldsymbol\delta} \|
    \widehat{\boldsymbol\Omega}_k\boldsymbol\delta
    -\widehat{\mathbf{g}}_k
    \|_1
    \text{ subject to } \begin{cases} l_i \le \widehat\vartheta_{k,i}+\delta_i \le u_i, \\
    |\delta_i| \le \max(1, |\widehat\vartheta_{k,i}|)\cdot\rho_k, \end{cases}
    \text{ for } i=1,\ldots, p.
    $$
    Here, $\|\cdot\|_1$ denotes the $\ell_1$-norm. The new candidate point is $\widetilde{\boldsymbol\vartheta}_k \gets \widehat{\boldsymbol\vartheta}_k+\boldsymbol\delta_k$.

    \State \label{alg:lstop} \emph{Convergence check}.
    If $L_k=\pp{NFit}{local}$ and $\|\widehat{\mathbf{g}}_k\|_{\mathbf{U}_k}^2 < p \cdot \pp{Tol}{local}$
    terminate the algorithm and go to the final step (\ref{end:local}).
    Here,  
    $\mathbf{U}_k=\widehat{\mathbf{J}}_k^{\mathrm{T}} 
    \widehat{\boldsymbol{\Sigma}}_k^{-1} \widehat{\mathbf{H}}_k
    \widehat{\boldsymbol{\Sigma}}_k^{-1} \widehat{\mathbf{J}}_k$ is an estimate
    of the dispersion matrix of $\widehat{\mathbf{g}}_k$.

    \State \label{alg:ladd}
    \emph{Sample around the candidate point}.
    Set $N_{k+1} \gets N_{k}+\pp{NAdd}{local}$ and draw 
    $\boldsymbol\vartheta_{N_k+1}, \ldots, \boldsymbol\vartheta_{N_{k+1}}$ uniformly from the ellipsoid
    $\{\boldsymbol\vartheta \in \Theta: 
        \|\boldsymbol\vartheta -\widetilde{\boldsymbol\vartheta}_k\|_{\boldsymbol\Omega_k^{-1}} 
        \le 1
     \}$. 
     Simulate the corresponding summary statistics $\mathbf{t}_{N_k+1}, \ldots, \mathbf{t}_{N_{k+1}}$.

    \State \label{alg:mok} \emph{Accept/reject step and adjust trust region}. 
    If
    $
    \sum_{i=N_k+1}^{N_{k+1}}
    \|\mathbf{t}_i-\widehat{\boldsymbol\tau}_k-
    \widehat{\mathbf{J}}_k(\boldsymbol\vartheta_i-\widetilde{\boldsymbol\vartheta}_k)
    \|_{\widehat{\boldsymbol\Sigma}_k}^2
    < q \cdot (N_{k+1}-N_k) \cdot \pp{Tol}{model},
    $
    accept the candidate point by setting
    $\widehat{\boldsymbol\vartheta}_{k+1} \gets
    \widetilde{\boldsymbol\vartheta}_k$ and expand the trust region setting
    $\rho_{k+1} \gets \min(2\rho_k, \pp{Rho}{max})$; otherwise, reject
    it by setting $\widehat{\boldsymbol\vartheta}_{k+1} \gets
    \widehat{\boldsymbol\vartheta}_k$ and shrink the trust region setting $\rho_{k+1} \gets \rho_k/4$.

    \State \label{alg:lupd}
    \emph{Update and iterate.} 
    Set $k \gets k+1$, update the neighborhood size $L_k \gets
    \min(\pp{NFit}{local}, L_{k-1}+\pp{NAdd}{local})$, and return to
    step \ref{alg:lest}.

    \State \label{end:local} \emph{Output}.
    Return $\widetilde{\boldsymbol\vartheta}_k$ as the final parameter estimate and $\widehat{\boldsymbol\Omega}_k^{-1}$ as its estimated covariance matrix.
\end{algorithmic}
\end{algorithm}

The local search phase, described in Algorithm~\ref{alg:local}, aims to
compute the final parameter estimate $\widehat{\boldsymbol{\vartheta}}$
by solving the estimating equation~\eqref{eqn:esteq} and provide an
estimate of its dispersion matrix. In the algorithm,
$\widehat{\boldsymbol{\vartheta}}_k$ denotes the current estimate of the
desired parameter $\widehat{\boldsymbol{\vartheta}}$. The procedure
begins with the best parameter vector identified during the global
exploration phase (Step~\ref{alg:linit}) and refines it through an
adaptation of the Fisher scoring method \citep{Osborne1992}. The
iterative update underlying this phase is given by
$$
\boldsymbol{\vartheta}_{\mathrm{new}} \gets
\boldsymbol{\vartheta}_{\mathrm{old}} +
\boldsymbol{\delta}_{\mathrm{new}},
$$
where the increment $\boldsymbol{\delta}_{\mathrm{new}}$ is the solution to
\begin{equation}
    \boldsymbol{\Omega}(\boldsymbol{\vartheta}_{\mathrm{old}})
    \boldsymbol{\delta}_{\mathrm{new}} =
    g(\boldsymbol{\vartheta}_{\mathrm{old}}), \label{eqn:delta}
\end{equation}
with $\boldsymbol{\Omega}(\cdot)$ and $g(\cdot)$ defined in equations~\eqref{eqn:asyvar} and~\eqref{eqn:grad}, respectively.

In our framework, both $\boldsymbol{\Omega}(\cdot)$ and $g(\cdot)$ are
unknown and must be approximated using Monte Carlo simulation. This
makes the direct application of standard Newton-type methods --- such as
Fisher scoring --- particularly prone to instability. These methods are
sensitive to step-size and curvature issues \citep{Nocedal2006}, and the
stochastic errors introduced by Monte Carlo estimation further
exacerbate this instability. To address this difficulty, the algorithm
incorporates a trust region mechanism \citep{Nocedal2006}.  This
approach enhances stability by applying a sequence of small corrections
to the initial solution, thereby mitigating the impact of inaccuracies
in the estimated iteration direction and preventing divergence through
successive refinements.

At each iteration, the quantities $\tau(\widehat{\boldsymbol{\vartheta}}_k)$,
$\mathbf{J}(\widehat{\boldsymbol{\vartheta}}_k)$, and
$\boldsymbol{\Sigma}(\widehat{\boldsymbol{\vartheta}}_k)$ --- and
consequently their functions $g(\widehat{\boldsymbol{\vartheta}}_k)$ and
$\boldsymbol{\Omega}(\widehat{\boldsymbol{\vartheta}}_k)$ --- are
estimated by fitting a linear model to the subset of simulated pairs
nearest to the current parameter vector (Step~\ref{alg:lest}). To reduce 
Monte Carlo noise further, the successive estimates of the Jacobian and 
the covariance matrix are smoothed using an exponentially weighted moving 
average.

The resulting local approximation is used to compute a new candidate
solution $\widetilde{\boldsymbol{\vartheta}}_k =
\widehat{\boldsymbol{\vartheta}}_k + \boldsymbol{\delta}_k$, where
$\boldsymbol{\delta}_k$ is determined by minimizing the
$\ell_1$-norm of the difference between the two sides of
equation~\eqref{eqn:delta}, subject to appropriate box constraints
(Step~\ref{alg:delta}). The parameter $\rho_k$ controls the size of the
trust region and is adjusted dynamically based on the accuracy of the
local linear model's approximation to the underlying quantities. In
particular, additional simulations are generated in the vicinity of the
new candidate point (Step~\ref{alg:ladd}). If the new simulations
confirm that the local model provides an adequate description, 
the candidate point is accepted and the trust
region is enlarged. Conversely, if the fit appears unreliable, the
update is rejected and the region is reduced, thereby restricting the
step size in the following iteration (Step~\ref{alg:mok}). 

The number of points in the neighborhood used for the local fit, denoted
by $L_k$ in Algorithm~\ref{alg:local}, starts at $\pp{N}{elite}$
(default: 100) and gradually increases --- by $\pp{NAdd}{local}$ points
per iteration (default: 10) --- until it reaches $\pp{NFit}{local}$
(default: 4000) (Step~\ref{alg:lupd}). The procedure terminates once
$g(\widehat{\boldsymbol{\vartheta}}_k)$ is sufficiently close to zero
(Step~\ref{alg:lstop}), but only after $L_k$ has reached
$\pp{NFit}{local}$.  The last requirement serves two purposes: it avoids
premature convergence and ensures that the standard errors, whose
estimation generally requires more simulations, are reasonably accurate.

The additional points generated during the local search also serve two
complementary purposes. First, they are used in future iterations to
refine the estimates of the mean $\boldsymbol{\tau}(\cdot)$ and the
covariance matrix $\boldsymbol{\Sigma}(\cdot)$ of the intermediate
statistics at the current estimate, suggesting that sampling should be 
performed as close as possible to the candidate point 
$\widetilde{\boldsymbol{\vartheta}}_k$. Second, these points contribute 
to improving the estimate of the Jacobian $\mathbf{J}(\cdot)$. As 
discussed, for example, by \citet{Cox2012}, this task is delicate, as two 
competing requirements must be balanced. Let $\widetilde{\boldsymbol{\vartheta}}_k +
\mathbf{h}$ be one of the newly sampled points. On the one hand,
$\mathbf{h}$ should be sufficiently small for the linear approximation
to hold; on the other hand, it must be large enough for the difference
$$ 
\boldsymbol{\tau}(\widetilde{\boldsymbol{\vartheta}}_k+\mathbf{h}) -
\boldsymbol{\tau}(\widetilde{\boldsymbol{\vartheta}}_k) 
\approx
\mathbf{J}(\widetilde{\boldsymbol{\vartheta}}_k)\mathbf{h} 
$$ 
to be significant relative to the simulation noise, as measured by the
dispersion matrix $\boldsymbol{\Sigma}(\boldsymbol{\vartheta})$. In
practice, this suggests that $\|
\mathbf{J}(\widetilde{\boldsymbol{\vartheta}}_k)\mathbf{h}
\|_{\boldsymbol{\Sigma}(\widetilde{\boldsymbol{\vartheta}}_k)}$ should
be small but non-negligible. This consideration underlies the choice of
the ellipsoid used to sample new points in Step~\ref{alg:ladd} of the
algorithm. Alternative sampling schemes were also explored, including
designs inspired by central composite design, commonly used within the
RSM framework~\citep{Myers2016}, or, as suggested by \citet{Cox2012} in
a related context, by the Plackett–Burman design. 
However, the differences proved negligible when the perturbations
$\mathbf{h}$ satisfied the size requirements discussed above, 
and, hence,  the simpler scheme described in Step~\ref{alg:ladd} 
was therefore retained.

\section{Case study: The Fowler's toad dataset}
\label{sec:toad}

\subsection{Data}

\begin{figure}
    \centering
    \begin{subfigure}{\textwidth}
    \caption{Observed displacements (meters) of 66 Fowler's toads from their
        initial refuge locations. White rectangles indicate missing data.}
   \includegraphics[width=\textwidth,
    height=0.6\textwidth]{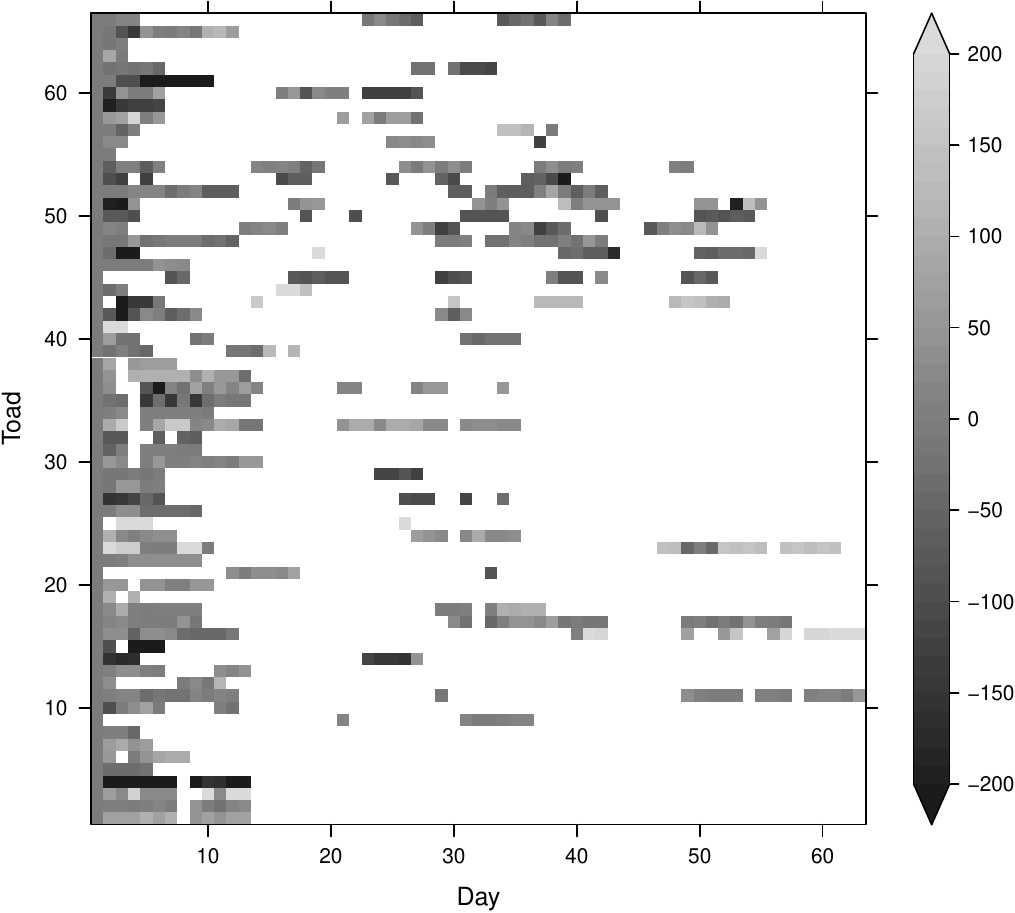}
    \label{fig:toad-data}
\end{subfigure}
\\
\begin{subfigure}[t]{0.45\textwidth}
    \caption{Parameter vectors sampled by \texttt{ifit}}
    \includegraphics[width=\textwidth,height=\textwidth]{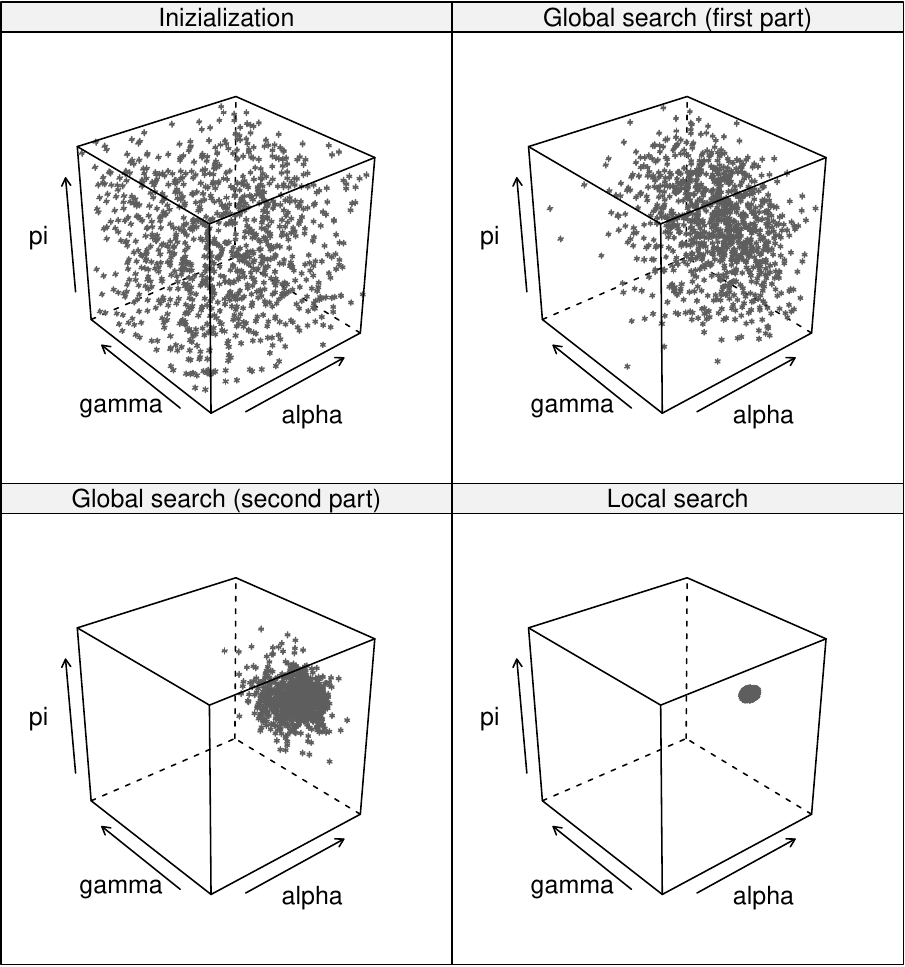}
    \label{fig:seqsampling}
\end{subfigure}  
\hfill
\begin{subfigure}[t]{0.45\textwidth}
    \caption{Standardized summary statistics}    
    \includegraphics[width=\textwidth, height=\textwidth]{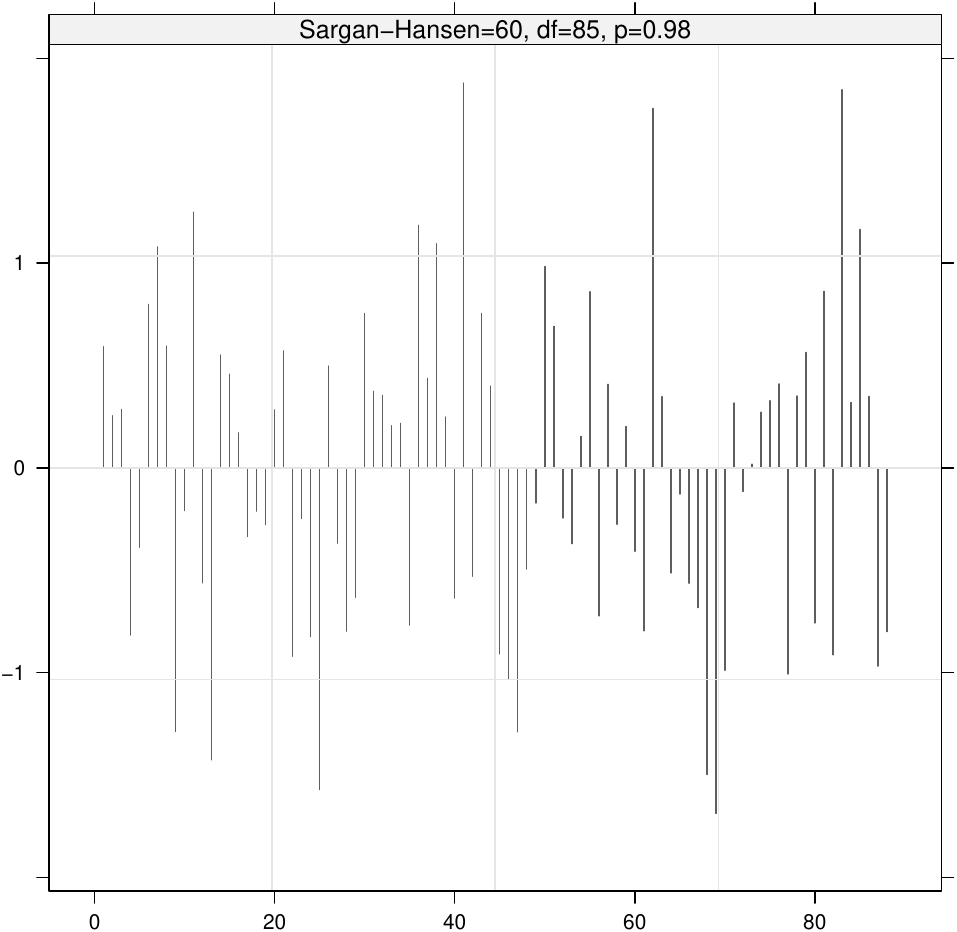}
    \label{fig:toad-diag}
\end{subfigure}  
\caption{Fowler's toad example}
\label{fig:toad-res}
\end{figure}

To illustrate the procedure, I analyzed the dataset presented by
\citet{Marchand2017}. The data describe the nocturnal movements of 66
Fowler's toads (\emph{Anaxyrus fowleri}) that were radio-tracked for 63
days along the north shore of Lake Erie, Ontario, Canada. Movement
occurs essentially along a single spatial dimension --- the one parallel to
the shoreline --- as motion in other directions is constrained by waves,
predators, and topography. For this reason, \citet{Marchand2017}
projected the recorded displacements onto a one-dimensional axis. The
data are shown in Figure~\ref{fig:toad-data}; note that more
than 80\% of the movement data are missing.

\subsection{Model}

The generative model considered is that named ``Model 1'' in
\citet{Marchand2017}.  This model assumes that toads leave their refuges
at night to forage and hide within sand dunes during the day. After the
$t$th nocturnal foraging phase, a toad is located at a displacement of
$\Delta_t$ meters from its previous refuge site. The displacements
$\Delta_t$ are assumed to be independent and identically distributed
realizations of a symmetric, zero-centered $\alpha$-stable random
variable with stability parameter $\alpha$ and scale parameter $\gamma$.
The toad then either returns to one of the previously visited sites
(with probability $\pi$) or remains at its current location (with
probability $1-\pi$). If the toad returns to a previous site, this is
selected uniformly at random. The model parameter vector is
$\boldsymbol{\vartheta} = (\alpha, \gamma, \pi)' \in (0, 2] \times [0,
+\infty) \times [0, 1]$.  When simulating from the model, the same
pattern of missing values observed in the real dataset is preserved.

\subsection{Summary statistics}

To summarize the data, similarly to  \citet{Marchand2017} and
\citet{An2022}, I adopted a statistic computed from the absolute
displacements at time lags of 1, 2, 4, and 8 days, classified as
``returns'' or ``non-returns'' depending on whether they are smaller
than $10$ meters. Specifically, for each lag, the statistic comprises:
the return frequency; and,  for the logarithms of the non-return
distances, the median and the differences between adjacent $p$th
quantiles ($p = 0.01, 0.05, 0.1, \ldots, 0.9, 0.95, 0.99$). 
This yields a 88-dimensional
summary vector, which is larger than the
12-dimensional statistic used by \citet{Marchand2017} and the
48-dimensional statistic used by \citet{An2022}, since it
is based on a more detailed description of the displacement
empirical distributions.

\subsection{Results}

\begin{table}
    \caption{Parameter estimates for the Fowler's toad case study.}
    \label{tab:toadest}
    \centering
    \begin{tabular}{llll}
        \toprule
        Parameter & Prior interval$^{\star}$ & Estimate & $95\%$
        Confidence/Credible interval\\
    \midrule
    \multicolumn{4}{c}{\texttt{ifit} (number of statistics: $88$; model simulations:
    $6800$)} \\
        $\alpha$ & $[0.01, 2]$ & $1.68$ & $[1.48,
        1.88]$\\
        $\gamma$ & $[0, 100]$ & $34.27$ & $[29.53, 39.02]$
        \\
        $\pi$ & $[0, 1]$ & $0.62$ & $[0.57, 0.68]$ \\[5pt]
        \multicolumn{4}{c}{\citet{Marchand2017} (number of statistics:
            12; model simulations: 10000)} \\
        $\alpha$ & $[1, 2]$ & $1.7$ & $[1.41,
        1.94]$\\
        $\gamma$ & $[10, 100]$ & $34$ & $[26, 42]$
        \\
        $\pi$ & $[0, 1]$ & $0.60$ & $[0.53, 0.65]$ \\[5pt]
        \multicolumn{4}{c}{\citet{An2022} (number of statistics:
            48; model simulations: 25000000)} \\
        $\alpha$ & $[1, 2]$ & $1.67$ & $[1.46,
        1.85]$\\
        $\gamma$ & $[0, 100]$ & $34$ & $[29, 39]$
        \\
        $\pi$ & $[0, 0.9]$ & $0.61$ & $[0.55, 0.67]$ \\[5pt]
        
\bottomrule
\multicolumn{4}{p{0.9\textwidth}}{\small$^\star$Box constraints for \texttt{ifit};
support of a uniform prior distribution for \citet{Marchand2017} and
\citet{An2022}.}
    \end{tabular}
\end{table}

Table~\ref{tab:toadest} reports the estimates (and the confidence
intervals based on asymptotic normality) obtained with
\texttt{ifit}, as well as those obtained for the same dataset using an
ABC rejection algorithm by \citet{Marchand2017} and an MCMC approach
based on the synthetic likelihood by \citet{An2022}.  
As can be seen, results are similar across methods and lead to the
same substantive conclusions. Notably, \texttt{ifit}, despite using a
higher-dimensional intermediate statistic and less restrictive prior
constraints on the parameters, achieves comparable results using fewer
simulations.

Figure~\ref{fig:seqsampling} illustrates the sequential sampling scheme.
Specifically, it shows the 6800 parameter vectors sampled in total,
divided into: (a)~the initial 1000 points used to initialize the
algorithm (simulations 1--1000); (b)~the next 1000 points sampled during
the first part of the global search (simulations 1001--2000); (c)~the
following 900 points sampled during the second part of the global search
(simulations 2001--2900); and (d)~the 3900 points sampled during the
local search stage (simulations 2901--6800). The figure highlights how
the explored region in the parameter space progressively shrinks,
concentrating around the final estimate.

Figure~\ref{fig:toad-diag} displays the summary statistics, standardized
using their estimated means and standard deviations computed from the
fitted model---see equation~(\ref{eqn:sscore}). The figure also reports
the value of the Sargan--Hansen test statistic (\ref{eqn:jtest}) and its
corresponding $p$-value. A high $p$-value and all standardized
statistics with absolute values below~2 indicate that the model
adequately describes the data---or at least the selected summary
statistics.

\section{Monte Carlo Simulations: Performance Evaluation and Comparisons}
\label{sec:num}

\subsection{Models}

\begin{figure}
    \centering
    \begin{subfigure}[t]{0.45\textwidth}
        \caption{\texttt{enzyme} ($\vartheta_1=0.5$, $\vartheta_2=2.5$,
        $\vartheta_3=1$)}
        \includegraphics[width=\textwidth]{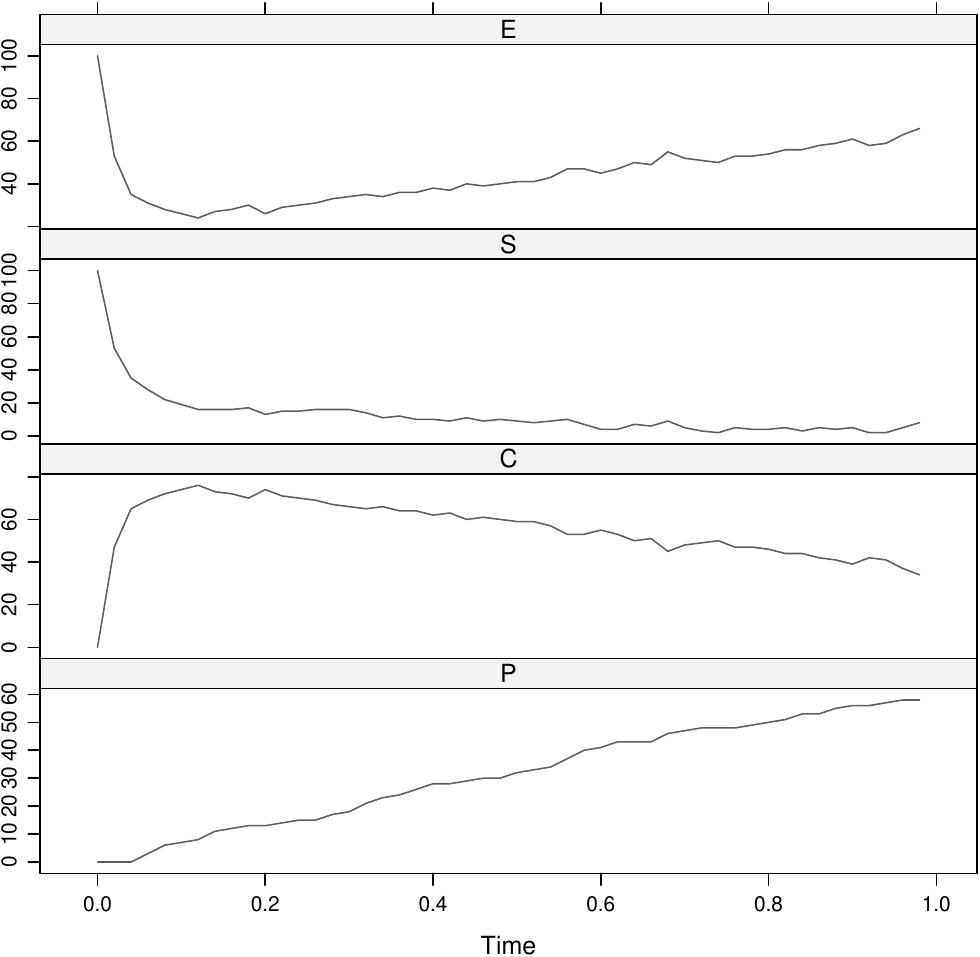}
        \label{fig:enzyme-ex}
    \end{subfigure} 
    \hfill
    \begin{subfigure}[t]{0.45\textwidth}
        \caption{\texttt{trait} ($\gamma=0.2$, $\mu=0.7$, $\sigma=0.1$,
        $\omega=0.7$)}
        \includegraphics[width=\textwidth]{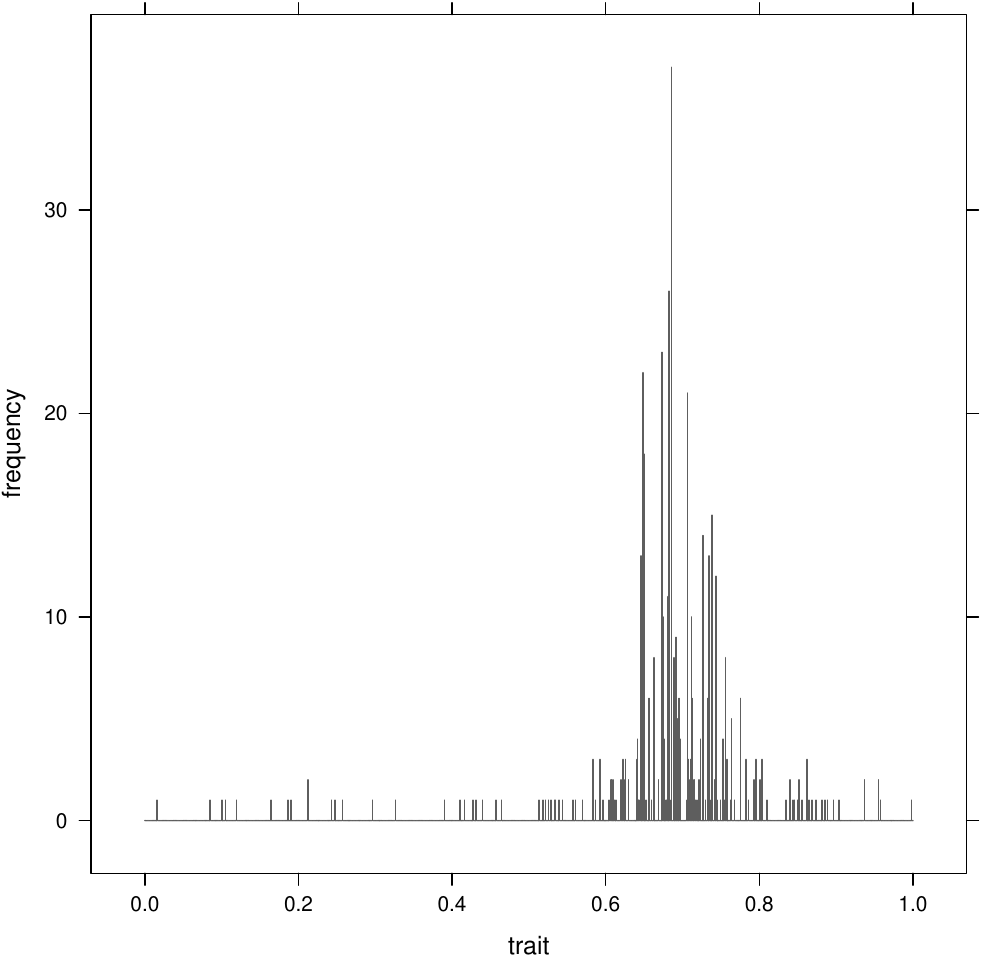}
        \label{fig:trait-ex}
    \end{subfigure} 
    \caption{Examples of simulated data}
    \label{fig:sim-ex}
\end{figure} 

The performance of \texttt{ifit} is evaluated for the following four models.

\emph{A logistic regression model including an intercept and three
covariates} (\texttt{logit}). Specifically,
$\mathbf{y}_{\mathrm{obs}} = (y_1, \ldots, y_n)^{T}$,
where $y_i$ are independent realizations of Bernoulli random variables
with success probability $\pi_i$ such that
$$
\log\frac{\pi_i}{1-\pi_i} = \vartheta_1 + \vartheta_2 x_i + \vartheta_3
z_i + \vartheta_4 w_i.
$$
In the simulations:
\begin{enumerate}[(i)]
    \item $n=100$; $x_i$ denotes a deterministic trend component, $x_i = (2i - n)/(n - 1)$;
    each vector $(z_i, w_i)^T$ is generated from a bivariate Gaussian distribution
    with $\text{E}(z_i)=\text{E}(w_i)=0$, $\text{var}(z_i)=1$, $\text{var}(w_i)=2$,
    and $\text{cov}(z_i, w_i)=1$;
    \item
    the ``true'' parameter vector is $\boldsymbol{\vartheta}_0 = (-1, 1, 0.5, -0.5)^{T}$;
    \item $\mathbf{t}_{\mathrm{obs}} = (\sum_i y_i, \sum_i x_i y_i,
        \sum_i z_i y_i, \sum_i w_i y_i)^T$; 
        that is, the summary statistics coincide with the sufficient
        statistics;
\item the box constraints are given by $l_i=-5$ and $u_i=5$ for $i = 1, \ldots, 4$.
\end{enumerate}

\emph{Remark.} This model, which has a tractable likelihood, is included
to benchmark \texttt{ifit} and its competitors against the maximum likelihood
estimator and Bayesian approaches that use the exact likelihood.

\emph{The stochastic Michaelis–Menten kinetic model}
(\texttt{enzyme}). The model describes the dynamics
of an enzyme $E$ that can bind to a substrate $S$ to form a complex $C$.
The complex $C$ then releases a product $P$ while regenerating the enzyme $E$
\citep{Wilkinson2018}.
The reactions are $E + S \rightarrow C$, $ C \rightarrow E + S$, and 
$C \rightarrow P + E$,
with rates that constitute the three model parameters.
Figure~\ref{fig:enzyme-ex} shows an example of simulated data.

In probabilistic terms, the integer-valued vector
$(E_t, S_t, C_t, P_t)^T$ evolves as a continuous-time Markov process such that:
\begin{align*}
&\Pr\{E_{t+\delta}=E_t-1, S_{t+\delta}=S_t-1, C_{t+\delta}=C_t+1, P_{t+\delta}=P_t
\mid E_t, S_t, C_t, P_t\} = \vartheta_1 E_t S_t \delta + o(\delta),  \\
&\Pr\{E_{t+\delta}=E_t+1, S_{t+\delta}=S_t+1, C_{t+\delta}=C_t-1, P_{t+\delta}=P_t
\mid E_t, S_t, C_t, P_t\} = \vartheta_2 C_t \delta + o(\delta),  \\
&\Pr\{E_{t+\delta}=E_t+1, S_{t+\delta}=S_t, C_{t+\delta}=C_t-1, P_{t+\delta}=P_t+1
\mid E_t, S_t, C_t, P_t\} = \vartheta_3 C_t \delta + o(\delta).
\end{align*}
Moreover, because the total amount of enzyme and substrate is conserved,
$E_t=E_0-C_t$ and $S_t=S_0-C_t-P_t$ at every $t$.
In the simulations:
\begin{enumerate}[(i)]
    \item
        the data consist of $E_t$, $S_t$, $C_t$, $P_t$ 
        sampled at 50 equally spaced points in the interval 
        $[0,1]$;
    \item the process is simulated using Gillespie's exact algorithm
        with initial conditions $E_0=100$, $S_0=100$, $C_0=0$, and $P_0=0$;
    \item the ``true'' parameter vector is $\boldsymbol{\vartheta}_0 = (0.5, 2.5, 1)^{T}$;
    \item the summary statistics are the coefficients of quadratic B-splines with a single knot at $0.2$, 
    fitted via least squares to the trajectories of $C_t$ and $P_t$;
    hence, the dimension of $\mathbf{t}_{\mathrm{obs}}$ is $q=8$; 
    \item the box constraints are given by $l_i=0$ and $u_i=50$ for $i =
        1, \ldots, 3$.
\end{enumerate}

\emph{The Jabot trait model} (\texttt{trait}).  The model describes the
distribution of a trait in a local community \citep{Jabot2010}.  The
trait takes values in $\{0, 0.001, \ldots, 0.999, 1\}$.  The competitive
ability of a species with trait $u$ is given by  $F(u) = 1 - \omega +
\omega \phi(u; \mu, \sigma)$, where $\omega$, $\mu$, and $\sigma$ are
parameters, and $\phi(u; \mu, \sigma)$ denotes the probability density
function of a normal random variable with mean $\mu$ and variance
$\sigma^2$.  The traits of the initial population are randomly drawn
with probability proportional to $F(u)$.  Then, at each time step, a
randomly drawn individual dies.  This individual is replaced either by
an immigrant (with probability $\gamma$) or by a descendant of an
existing individual (with probability $1 - \gamma$).  In the immigration
case, the new individual's trait is drawn with probability proportional
to $F(u)$.  In the reproduction case, the probability that the trait of
the new individual is $u$ is proportional to the abundance of $u$ in the
local community times $F(u)$.  The parameter vector is
$\boldsymbol{\vartheta} = (\gamma, \mu, \sigma, \omega)^{T}$.
Figure~\ref{fig:trait-ex} shows an example of simulated data.  
In the simulations, 
\begin{enumerate}[(i)] 
    \item 
        the population size and the
    number of time steps are set to 500 and 5000, respectively;
    \item 
        the ``true'' parameter vector is $\boldsymbol{\vartheta}_0 = (0.2, 0.7, 0.1, 0.7)^T$; 
    \item
        the summary statistics include the species richness (i.e., the
        number of distinct traits in the final population), the Gini
        index, and the $p$-quantiles of the trait for $p = 0.01$,
        $0.05$, $0.10, \ldots$, $0.90$, $0.95$, $0.99$; hence, the
        dimension of $\boldsymbol{t}_{\mathrm{obs}}$ is $q=21$; 
    \item 
        the box constraints are given by $l_i=0$ and $u_i=1$ for $i=1,\ldots, 4$.  
\end{enumerate}  

\emph{Remark.} The parametrization adopted here differs from that
originally proposed by \citet{Jabot2010}.  Specifically, Jabot defines
$\gamma = J / (J + \text{(local population size)} - 1)$ and $F(u) = 1 +
2A\pi\sigma \phi(u; \mu, \sigma)$, where $J$ and $A$ replace $\gamma$
and $\omega$, respectively.  However, the parametrization adopted here
facilitates specifying sensible parameter bounds.

\emph{Toad movement model} (\texttt{toad}).
The model, along with its summary statistics and prior restrictions on the parameters,
is described in Section~\ref{sec:toad}. In the simulations,
the ``true'' parameter vector is $\boldsymbol{\vartheta}_0=(\alpha_0, \gamma_0, \pi_0)^T=(1.7, 35, 0.6)$.
For this example, the dimension of $\mathbf{t}_{\mathrm{obs}}$ is
$q=88$.

\subsection{Alternative estimators}

I selected three possible competitors for \texttt{ifit} whose implementation was
available on the \emph{Comprehensive R Archive Network} (CRAN). 
For the Bayesian estimators, a uniform prior over the hypercube
$\boldsymbol\Theta$ defined by the box constraints used for \texttt{ifit} is employed.

The first competitor, which I denote as \texttt{ssfit}
\citep[see the package][]{ssfit-package}, is the estimator of \citet{Cox2012}. It
approximates $\boldsymbol\tau(\cdot)$, $\boldsymbol\Sigma(\cdot)$ and
related quantities by simulating multiple times for the same parameter
values. As suggested by \citet{Cox2012}, the procedure can be iterated
until the stopping rule proposed by the authors is satisfied. It
therefore corresponds to a classic local optimization algorithm.  I
initialize at $(l_i+u_i)/2$ for each parameter, use $4000$ simulations
per iteration (matching the neighborhood size of \texttt{ifit}), and set
the step size used for estimating the Jacobian matrix to $0.05$ for all
parameters.  Moreover, in our simulations, 
we allowed for a maximum of 25 iterations (i.e., $100{,}000$ simulations
from the model).

The second competitor, denoted as \texttt{apmc}, is the ABC algorithm
proposed by \citet{Lenormand2013} implemented in the \texttt{EasyABC}
package \citep{EasyABC-package}.  Like \texttt{ifit}, the algorithm aims
to minimize the required simulations by using sequential Monte Carlo
sampling, and, among the sequential algorithms implemented in
\texttt{EasyABC}, it is the only one that does not require
model-specific information (in particular, regarding the expected
distance from $\mathbf{t}_{\mathrm{obs}}$).  Indeed, when using the
default settings, the only tuning parameter is the number of posterior
samples, set here to $1000$.

The third competitor, denoted as \texttt{abcrf}, is the random-forest
ABC estimator proposed by \citet{Raynal2019} and implemented in the
\texttt{abcrf} package \citep{abcrf-package}. It shares with
\texttt{ifit} a focused objective: to provide a point estimate of the
parameters accompanied by a reasonable assessment of the standard
errors. The only tuning parameter is the number of simulations, set here
to $50{,}000$. While \texttt{abcrf} allows the reuse of model
simulations to estimate parameters for different datasets, this feature
was not exploited to ensure homogeneity with the other estimators; the
$50{,}000$ model simulations were regenerated for each replication in the
simulation study.

For the \texttt{logit} model, where the likelihood is tractable, we also
compare against the maximum likelihood estimator (\texttt{mle}) and a
Bayesian estimator (\texttt{jags}) implemented via MCMC using the
\texttt{rjags} package \citep{rjags-package} 
and the underlying JAGS library \citep{JAGS-manual}.

\subsection{Results}

\begin{table}
    \caption{Summary of the simulation results}
    \label{tab:synth}
\centering 
\begin{tabular*}{0.9\linewidth}{@{\extracolsep{\fill}}rrrrrrr}
  \toprule
 & \texttt{ifit} & \texttt{ssfit} & \texttt{apmc} & \texttt{abcrf} &
    \texttt{mle} & \texttt{jags} \\ 
  \midrule
  \multicolumn{7}{c}{\texttt{logit} (number of parameters: 4; number of
  summary statistics: 4)}\\
AMS & 8818 & 35524 & 44340 & 50000 &  &  \\ 
AARE & 0.403 & 0.430 & 1.115 & 1.228 & 0.400 & 0.437 \\ 
  \multicolumn{7}{c}{\texttt{enzyme} (number of parameters: 3; number of
  summary statistics: 8)}\\
AMS & 9449 & 97853 & 78533 & 50000 &  &  \\ 
AARE & 0.140 & 2.534 & 1.667 & 5.729 &  &  \\ 
  \multicolumn{7}{c}{\texttt{trait} (number of parameters: 4; number of
  summary statistics: 21)}\\
AMS & 7104 & 49316 & 47792 & 50000 &  &  \\ 
AARE & 0.062 & 0.117 & 0.092 & 0.093 &  &  \\ 
  \multicolumn{7}{c}{\texttt{toad} (number of parameters: 3; number of
  summary statistics: 88)}\\
AMS & 6761 & 85688 & 33643 & 50000 &  &  \\ 
AARE & 0.056 & 0.182 & 0.063 & 0.054 &  &  \\ 
   \bottomrule
\end{tabular*}
\end{table}

\begin{figure}
    \centering
    \includegraphics[width=0.50\textwidth,
    height=0.50\textwidth]{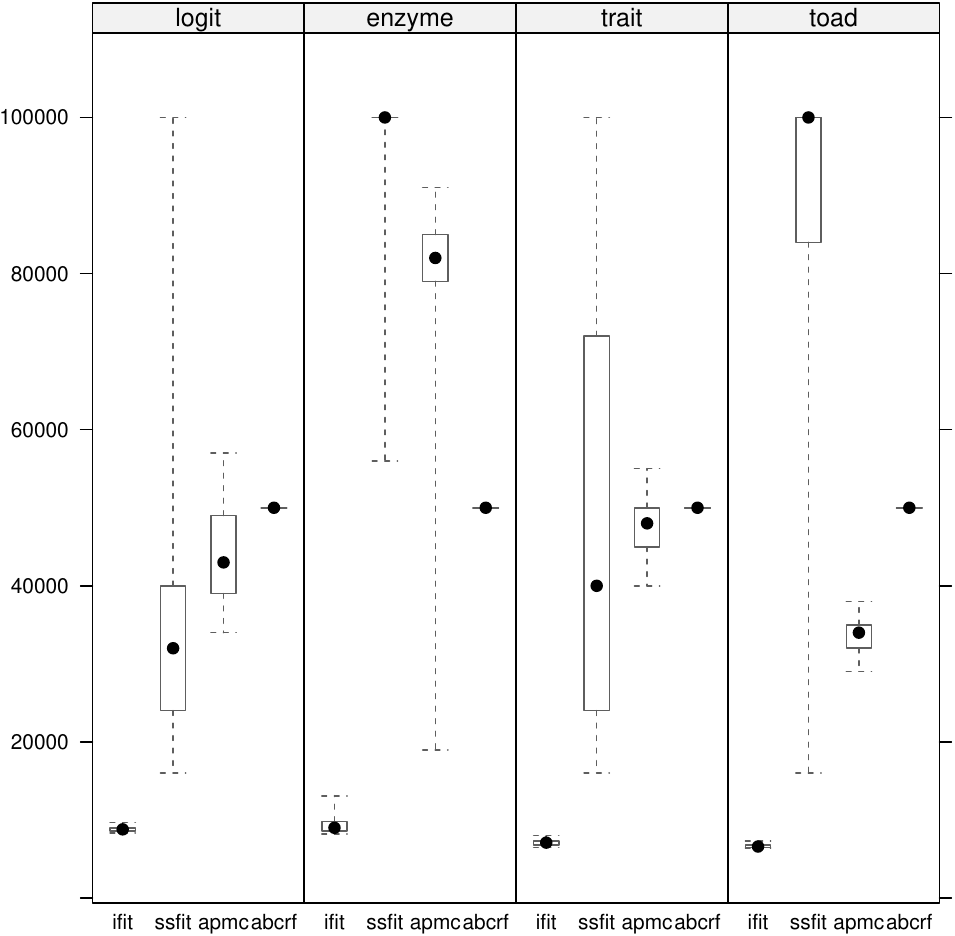}
    \caption{Number of model simulations. 
    The boxplots show the empirical $0.025$, $0.25$, $0.5$, $0.75$, and $0.975$ quantiles
    obtained from 1000 Monte Carlo replications.}
\label{fig:nsim}
\end{figure}  

\begin{figure}
    \centering
    \includegraphics[width=0.45\textwidth]{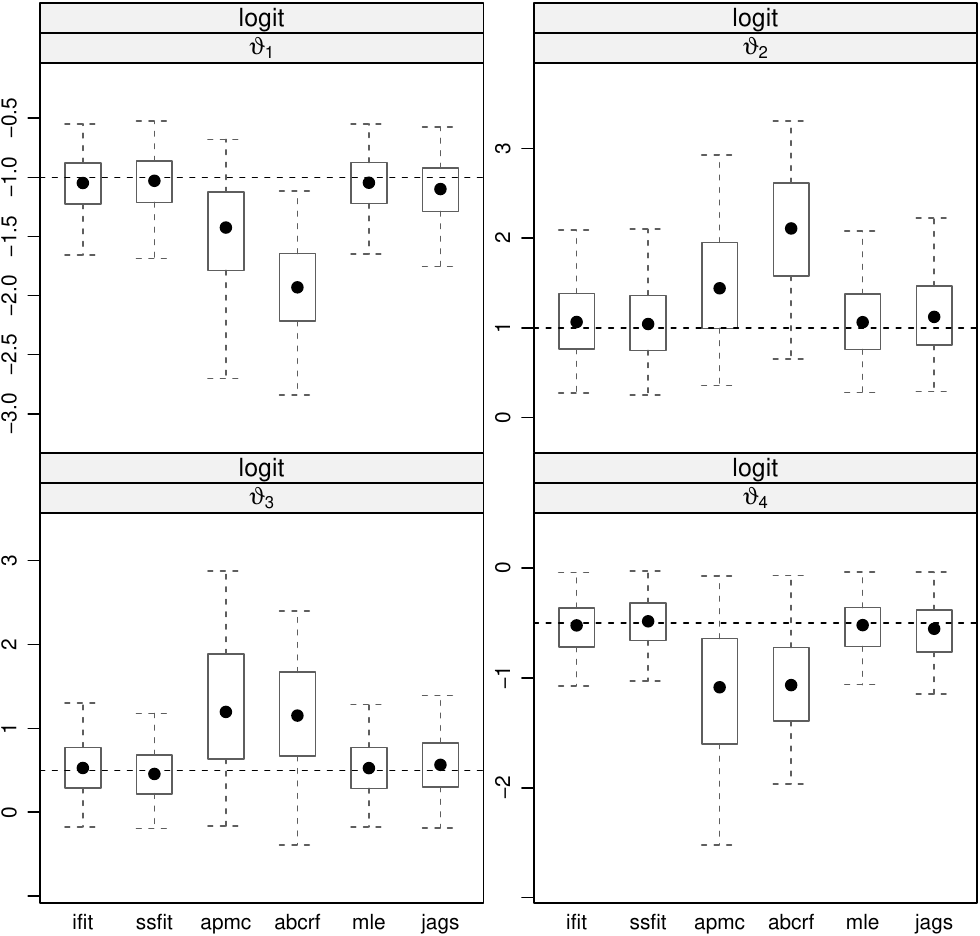}
    \includegraphics[width=0.45\textwidth]{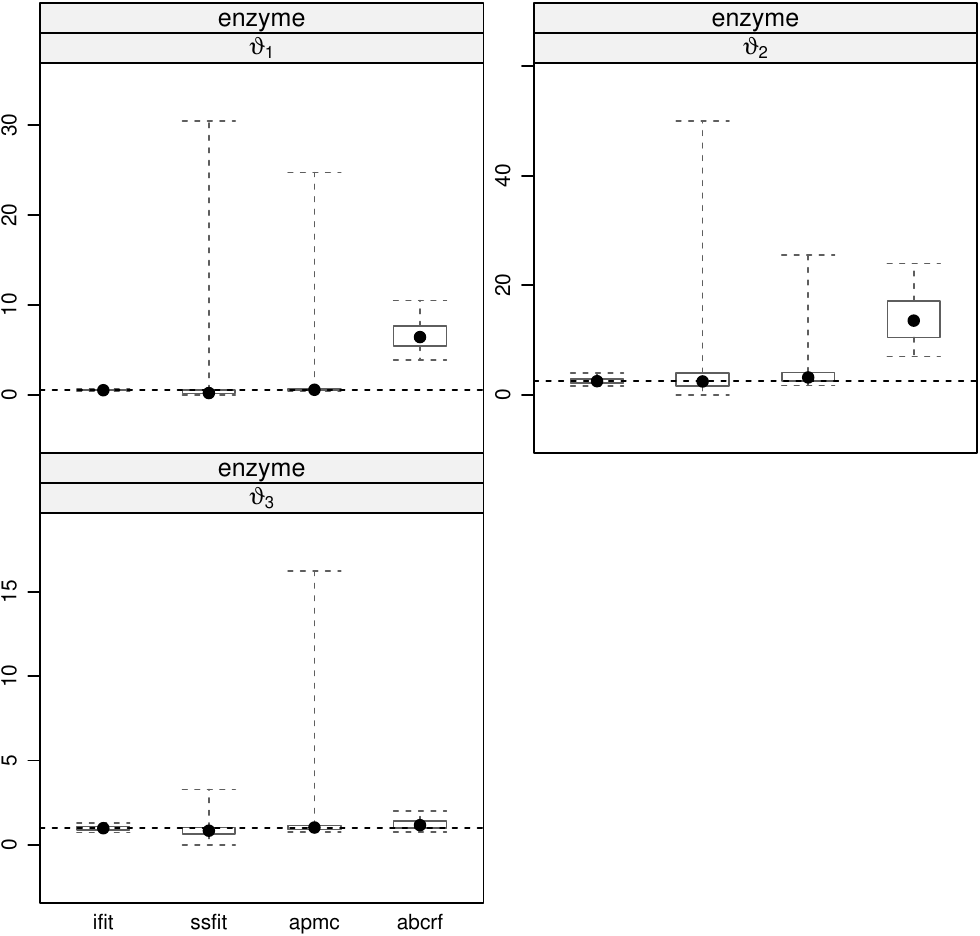}
    \\
    \includegraphics[width=0.45\textwidth]{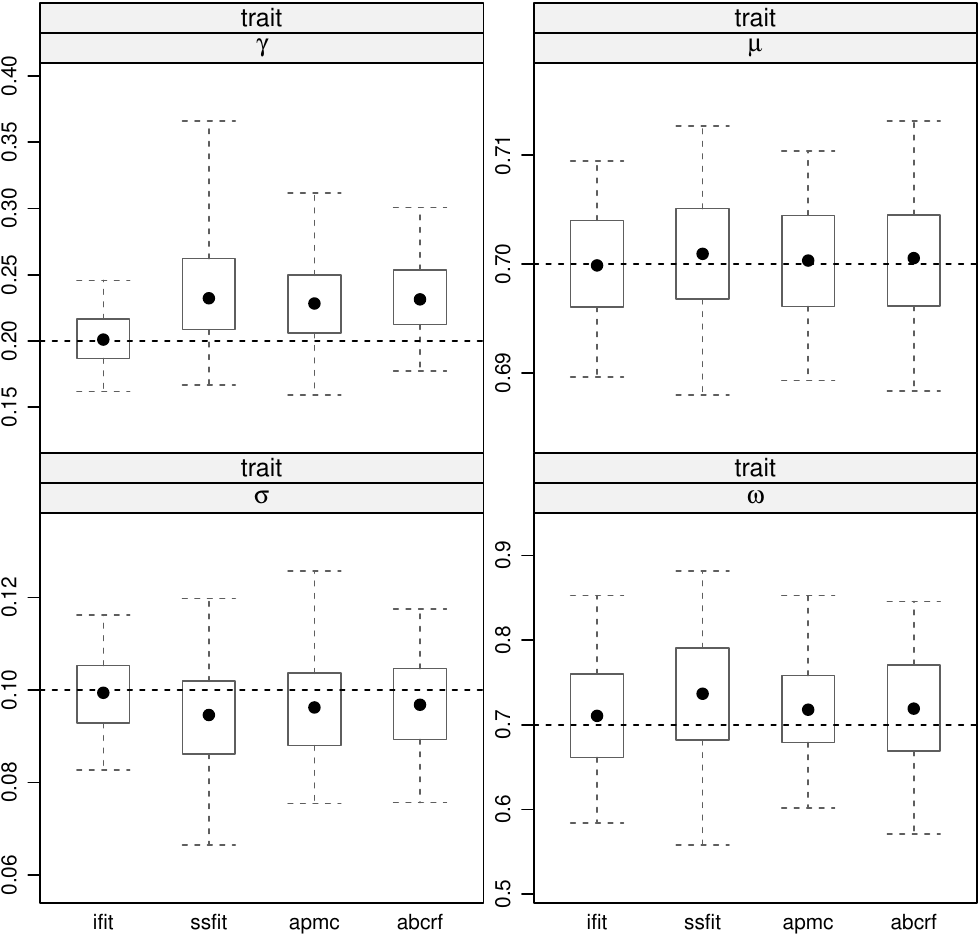}
    \includegraphics[width=0.45\textwidth]{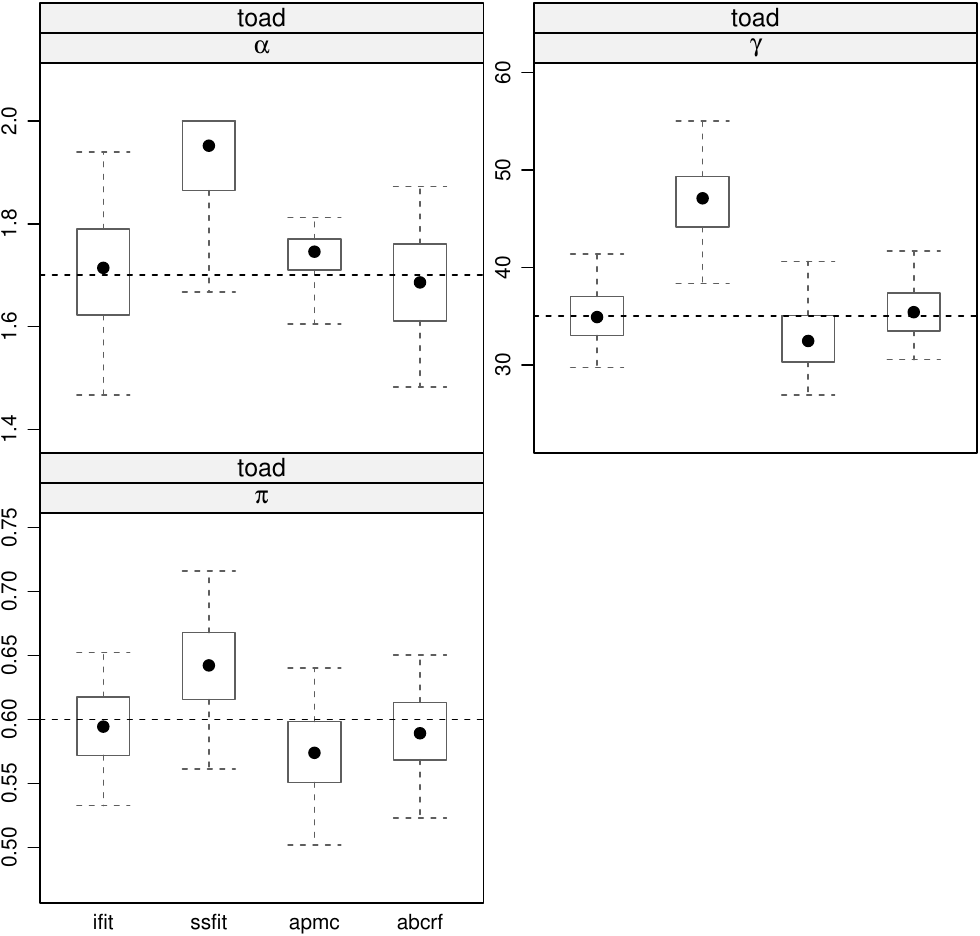}
    \caption{Empirical sampling distributions of the estimators. 
    The boxplots show the empirical $0.025$, $0.25$, $0.5$, $0.75$, and $0.975$ quantiles
    obtained from 1000 Monte Carlo replications.}
\label{fig:estimates}
\end{figure}  

For each model, $B = 1000$ synthetic datasets were generated, and the corresponding
estimators were computed.
Table~\ref{tab:synth} summarizes the results using the
average number of model simulations (AMS) and the average absolute
relative error (AARE) indices, defined as follows:
\begin{align*}
    \mathrm{AMS} &= \dfrac{\text{total number of model simulations performed across all $B$ replications}}{B},  \\
    \mathrm{AARE} &= \dfrac{1}{p \times B} \sum_{i=1}^p \sum_{j=1}^B
           \left|\dfrac{%
               \text{(estimate of parameter $i$ for dataset $j$)} -
               \text{(true value of parameter $i$)}
            }{%
               \text{(true value of parameter $i$)}
            }
            \right|.
\end{align*}   

Figures~\ref{fig:nsim} and~\ref{fig:estimates} respectively show the empirical
distributions of the number of model simulations and of the parameter
estimates obtained in the four test cases.
For the \texttt{enzyme} model, the reported results for \texttt{ssfit}
are based on 678 replications only, since the algorithm failed
in the remaining cases because of matrix singularities.

The results are clear: although \texttt{ifit} required substantially fewer
model simulations than its competitors, it was either the best performer or among
the top performers in terms of estimation accuracy across all four scenarios.
Conversely, the three ``likelihood-free'' competitors (\texttt{ssfit}, \texttt{apmc}, and
\texttt{abcrf}) not only required greater computational resources but also
exhibited poorer estimation accurancy in one or more cases.

Naturally, these results depend not only on the models considered but also
on the fact that the same default configuration was used for all four models.
For example, the poor performance of \texttt{ssfit} on the \texttt{enzyme} model
might improve if initialization were less ``neutral'' and based on
preliminary estimates obtained from simulated datasets.
Similarly, the two ABC algorithms might perform better under more informative
prior distributions.
However, the goal of this experiment was precisely to compare performance
across different scenarios using a uniform, minimally informed setup—one
that avoids user-dependent tuning or the need for prior knowledge.
In this context, the advantages of \texttt{ifit} clearly emerge.

It is also possible that the competitors’ accuracy could increase by
increasing the number of model simulations --- for example, by adopting a more
stringent stopping rule for \texttt{apmc} than the default proposed
by the authors and implemented in the \texttt{EasyABC} package, or by increasing the
number of simulated datasets used to train the random forests for
\texttt{abcrf}. However, this would further widen the computational
cost gap in favor of \texttt{ifit}.

\begin{figure}
    \centering
    \begin{subfigure}{0.45\textwidth}
        \caption{\texttt{ifit} vs \texttt{mle}}
        \includegraphics[width=\textwidth]{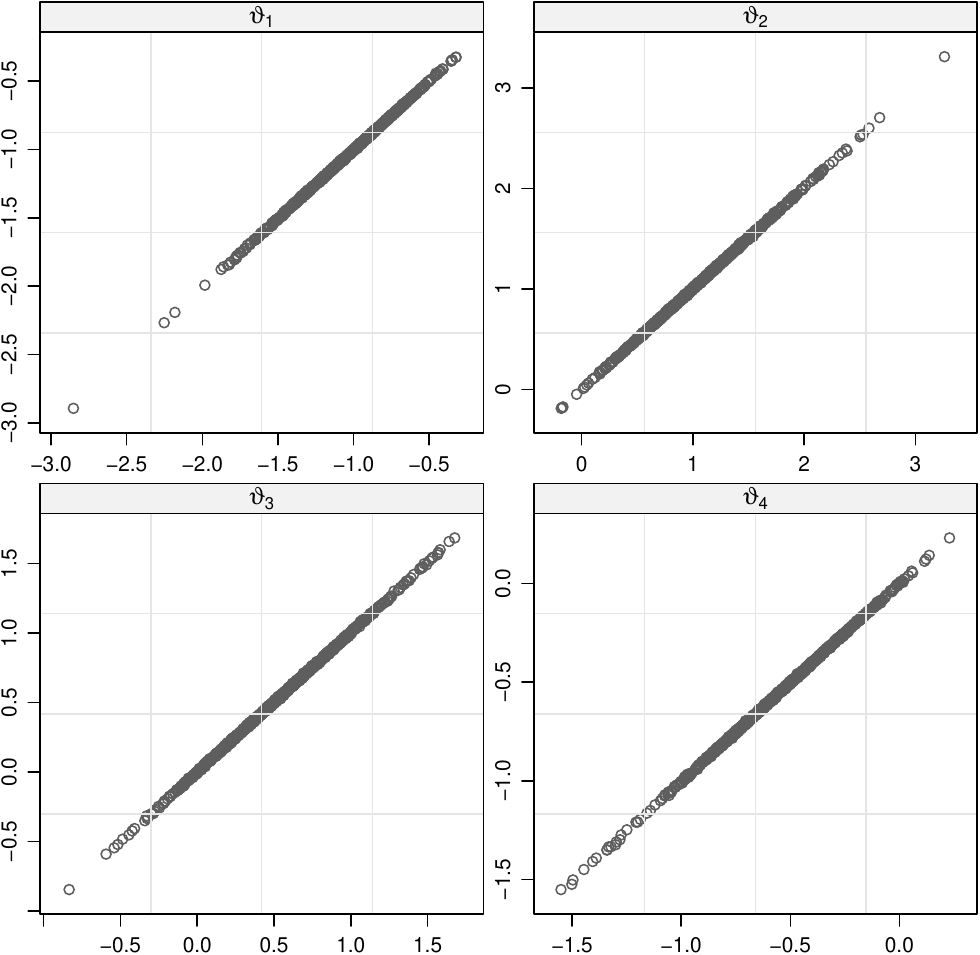}
    \end{subfigure}  
    \hfill
    \begin{subfigure}{0.45\textwidth}
        \caption{\texttt{ssfit} vs \texttt{mle}}
        \includegraphics[width=\textwidth]{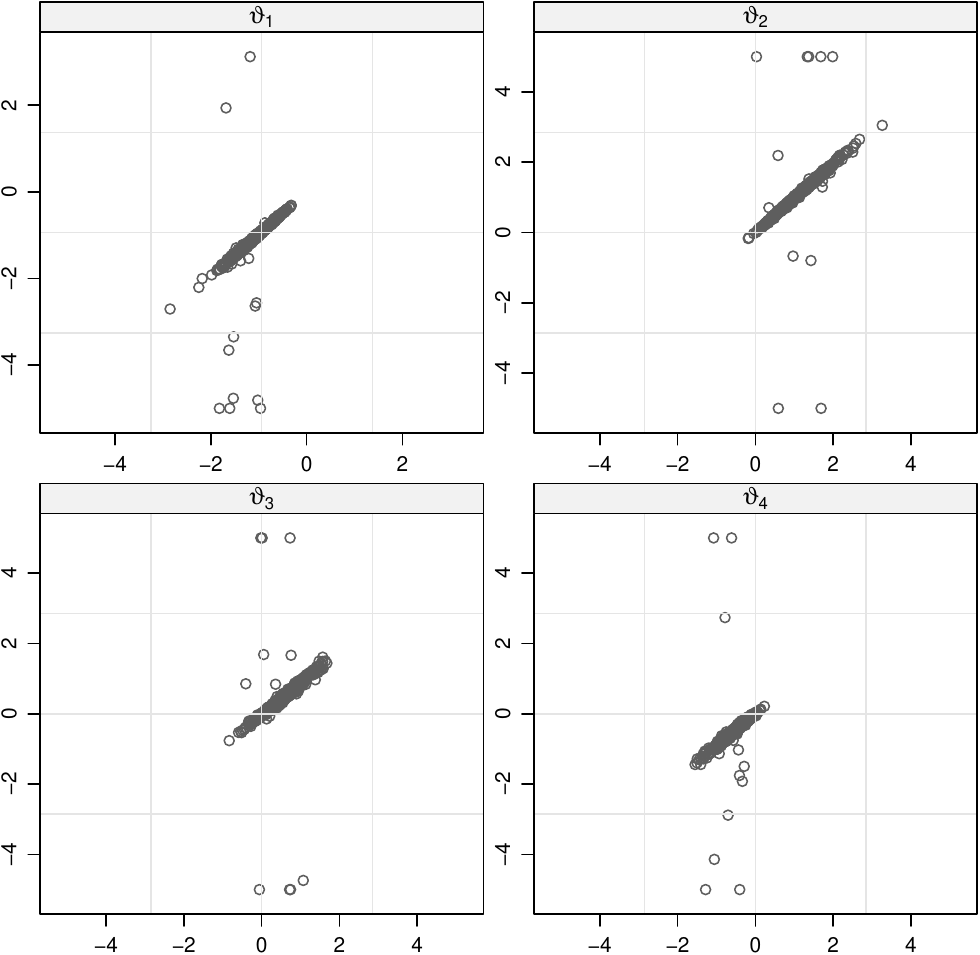}
    \end{subfigure}  
    \\[10pt]
    \begin{subfigure}{0.45\textwidth}
        \caption{\texttt{apmc} vs \texttt{jags}}
        \includegraphics[width=\textwidth]{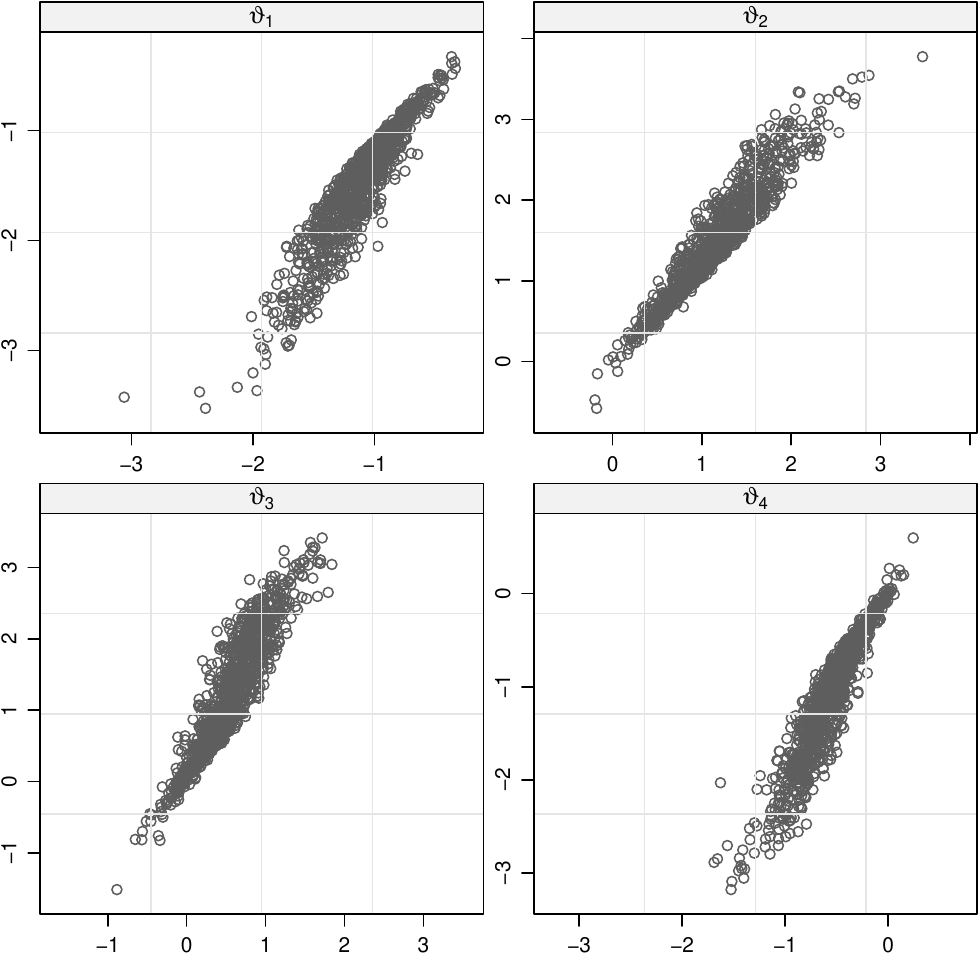}
    \end{subfigure}  
    \hfill
    \begin{subfigure}{0.45\textwidth}
        \caption{\texttt{abcrf} vs \texttt{jags}}
        \includegraphics[width=\textwidth]{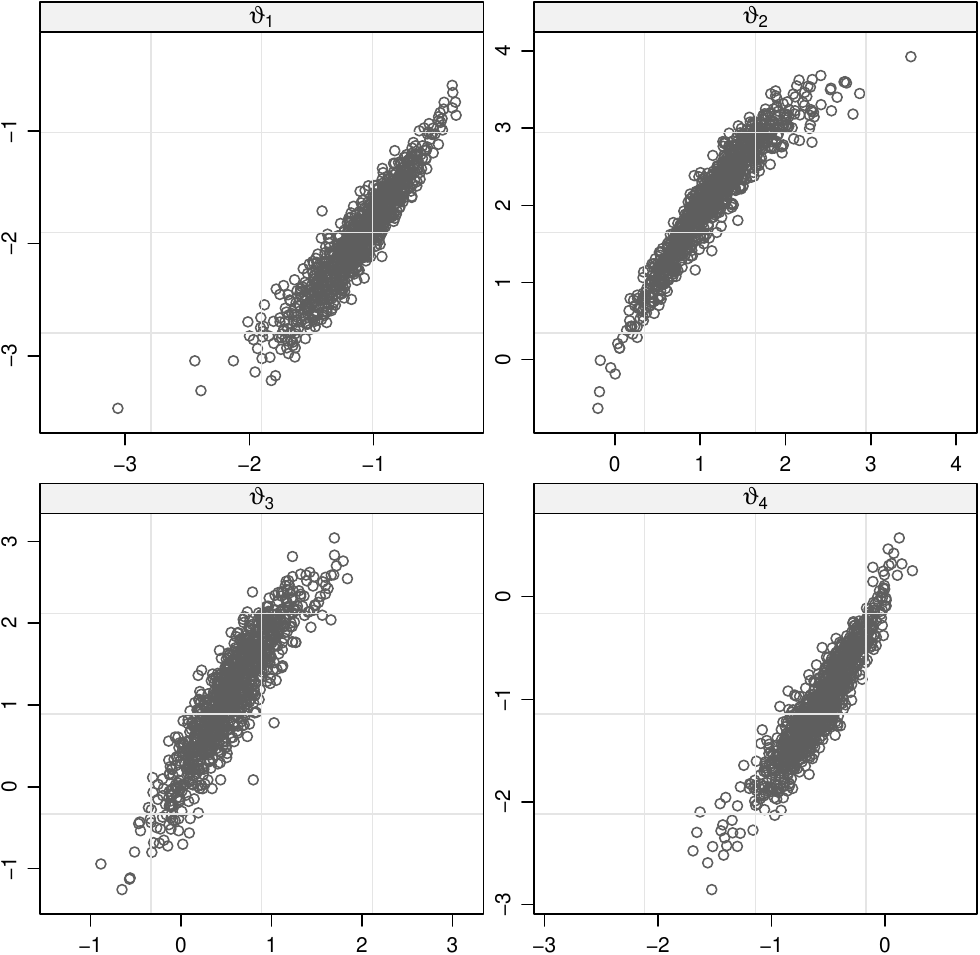}
    \end{subfigure}  
    \caption{``Likelihood-based'' versus ``likelihood-free'' estimators
    for the \texttt{logit} model.}
    \label{fig:based-free}
\end{figure}   

Figure~\ref{fig:based-free} shows, for the \texttt{logit} model, a
comparison between estimators that exploit knowledge of the likelihood
and those that do not.
Note that \texttt{ifit} reproduces the maximum likelihood estimator
in each replication with negligible  error, which is, after all, its
objective: the estimating function~\eqref{eqn:grad} coincides
with the likelihood score for the \texttt{logit} model, since the model belongs
to the exponential family and the chosen summary statistics are sufficient.
Conversely, this is not the case for the other three competitors.
In particular, \texttt{ssfit} occasionally deviates substantially
from \texttt{mle}, suggesting convergence issues.
Meanwhile, both \texttt{apmc} and \texttt{abcrf} only approximate
the results of their Bayesian ``likelihood-based''
counterpart, \texttt{jags}, with systematic bias and high
variability.

\begin{table}
    \centering
    \caption{Evaluation of \texttt{ifit} standard error estimates}
    \label{tab:se}
    \begin{tabular*}{0.8\linewidth}{@{\extracolsep{\fill}}crrrrrrrr}
        \toprule
        & \multicolumn{4}{c}{\texttt{logit}} 
        && \multicolumn{3}{c}{\texttt{enzyme}} \\
        & $\vartheta_1$ & $\vartheta_2$ & $\vartheta_3$ & $\vartheta_4$
        &&
        $\vartheta_1$ & $\vartheta_2$ & $\vartheta_3$ \\
        \cmidrule(r){2-5}
        \cmidrule(r){7-9}
        $se$ & 0.286 & 0.461 & 0.375 & 0.267 &  & 0.076 & 0.587 & 0.144 \\ 
        ave($\widehat{se}$) & 0.258 & 0.439 & 0.360 & 0.260 &  & 0.073 & 0.556 & 0.133 \\ 
        sd($\widehat{se}$)  & 0.035 & 0.052 & 0.046 & 0.037 &  & 0.015 & 0.117 & 0.015 \\ 
        \midrule
        & \multicolumn{4}{c}{\texttt{trait}} 
        && \multicolumn{3}{c}{\texttt{toad}} \\
        & $\gamma$ & $\mu$ & $\sigma$ & $\omega$
        &&
        $\alpha$ & $\gamma$ & $\pi$ \\
        \cmidrule(r){2-5}
        \cmidrule(r){7-9}
        se & 0.022 & 0.005 & 0.009 & 0.071 &  & 0.121 & 2.933 & 0.032 \\ 
        ave($\widehat{se}$) & 0.022 & 0.005 & 0.009 & 0.072 &  & 0.101 & 2.517 & 0.028 \\ 
        sd($\widehat{se}$)  & 0.004 & 0.001 & 0.001 & 0.008 &  & 0.009 & 0.287 & 0.002 \\ 
        \bottomrule
    \end{tabular*}  
\end{table}

Table~\ref{tab:se} assesses the accuracy of the standard errors of the
parameters estimates provided by \texttt{ifit}. Specifically, let $\hat{\vartheta}_{i,j}$ and
$\widehat{se}_{i,j}$ denote, respectively, the estimate of the $i$-th parameter and its
estimated standard error obtained in the $j$-th replication. 
For each parameter $i$, the table reports the empirical standard deviation of
$\hat{\vartheta}_{i,j}$ ($se$), together with the mean (ave($\widehat{se}$)) and the
standard deviation (sd($\widehat{se}$)) of the estimated standard errors
$\widehat{se}_{i,j}$. The first quantity represents the true sampling
variability,
whereas the latter two describe how well it is captured by \texttt{ifit}.
The results show that the standard errors produced by \texttt{ifit} are quite
reasonable, although, as might be expected, they tend to slightly underestimate
the true sampling variability.

\begin{table}
    \caption{Importance of Monte Carlo errors in \texttt{ifit}}
    \label{tab:rep}
\centering
    \begin{tabular*}{0.8\linewidth}{@{\extracolsep{\fill}}rrrrrrrr}
    \toprule
         \multicolumn{4}{c}{\texttt{logit}} 
        && \multicolumn{3}{c}{\texttt{enzyme}} \\
         $\vartheta_1$ & $\vartheta_2$ & $\vartheta_3$ & $\vartheta_4$
                       &&
        $\vartheta_1$ & $\vartheta_2$ & $\vartheta_3$ \\
        \cmidrule(r){1-4}
        \cmidrule(r){6-8}
        \\
        0.0003 & 0.0002 & 0.0002 & 0.0002 && 
        0.0081 & 0.0108 & 0.0083 \\  
\midrule
         \multicolumn{4}{c}{\texttt{trait}} 
       && \multicolumn{3}{c}{\texttt{toad}} \\
         $\gamma$ & $\mu$ & $\sigma$ & $\omega$
                  &&
        $\alpha$ & $\gamma$ & $\pi$ \\
        \cmidrule(r){1-4}
        \cmidrule(r){6-8}
        \\
        0.0250 & 0.0081 & 0.0282 & 0.0571 &&
  0.0867 & 0.1245 & 0.0828  \\ 
\bottomrule
\end{tabular*}
\end{table}  

In simulation-based estimation, it is essential that the additional
Monte Carlo error remains small compared with the total estimation
error. Only in this case can the results be considered reproducible.  To
assess this aspect, for each of the four models, I generated 100
additional datasets and estimated the parameters with \texttt{ifit} ten
times for each dataset. Then, I computed the standard ANOVA
decomposition of the total sum of squares into between- and
within-groups, using the 100 datasets as grouping factor.
Table~\ref{tab:rep} reports the ratios between the within-group sums of
squares --- which measure the variability due to Monte Carlo
error --- and the total sums of squares --- which represent the overall
variability of the estimator.  As desired, these ratios are consistently
small, indicating that simulation noise has only a minor impact on the
results.

\section{Conclusions}
\label{sec:conc}

The starting point of this work was to explore whether combining a
gradual sequential learning scheme with local smoothing techniques could
lead to a reasonably efficient and broadly applicable algorithm --- at least
as effective as the best procedures proposed within the ABC
framework --- for estimating simulation-based models with intractable
likelihoods, particularly under vague or weakly informative prior
knowledge.

The results obtained so far are encouraging, suggesting that the
proposed approach deserves further development. In particular, two
research directions appear especially promising:

\begin{enumerate} 
\item \emph{High $p$ and/or $q$ scenarios.} 
    The numbers of parameters $p$ and summary statistics $q$ affect both
    the efficiency and the number of model simulations required to
    estimate the Jacobian matrix $\mathbf{J}(\boldsymbol\vartheta)$, of
    dimension $q \times p$, and the variance matrix
    $\boldsymbol\Sigma(\boldsymbol\vartheta)$, of dimension $q \times
    q$.

    In addition to the models discussed in this paper, I have
    successfully used \texttt{ifit}, with the suggested default
    settings, to estimate models with up to $p = 12$ parameters. This
    seems to cover a wide range of practical situations where
    simulation-based model fitting techniques have been applied.
    However, it remains to be investigated whether and how \texttt{ifit}
    can be extended to handle even larger parameter spaces. For
    instance, one possibility would be to estimate the rows of the
    Jacobian via local regression methods combined with some form of
    variable selection.

    Regarding $q$, it is worth noting that even in its current version,
    \texttt{ifit} can handle fairly large values of $q$
    --- for example, 88
    summary statistics were used in the toad movement example.
    A natural extension for efficiently dealing with an
    even larger number of summary statistics would be to adopt
    high-dimensional estimators of the covariance matrix
    \citep[e.g.,][]{Warton2008, Cai2011, Pourahmadi2013}.
    Alternatively, one could reduce the set of summary statistics by
    selecting the most informative ones --- or even redefining them, as 
    suggested in the ABC framework by \citet{Fearnhead2012} ---
    at the end of the global search stage.

    \item \emph{Model selection.} 
        Model selection remains a central
        issue in many applications. A promising avenue for future
        research would be to use the final outputs of \texttt{ifit}
        --- in particular, the estimated mean vector and covariance matrix of
        the summary statistics --- to approximate the synthetic likelihood
        \citep{Wood2010}. This approximation could then serve as a basis
        for computing information criteria such as AIC, BIC, or related
        measures for model comparison.  
\end{enumerate}

\section*{ACKNOWLEDGMENT}

This work was completed during a months-long hospital stay that ended
with a delicate heart surgery. I am profoundly grateful to the hospital
staff — from care assistants and nurses to residents and doctors — for
their professionalism, their smiles, the light-hearted banter, the
late-night ice creams, and their steadfast support, as well as to my
many roommates for the camaraderie.

\bibliographystyle{abbrvnat}
\bibliography{ifit}

\end{document}